\documentclass[pop, aip, amsmath, amssymb, floatfix, preprint]{revtex4-1}
\usepackage{graphicx}
\usepackage{dcolumn}
\usepackage{bm}

\usepackage[utf8]{inputenc}
\usepackage[T1]{fontenc}
\usepackage{mathptmx}

\usepackage{amssymb,amsfonts,amsmath}
\usepackage{etoolbox}

\usepackage{booktabs}
\usepackage{soul}
\usepackage{siunitx}
\usepackage{xcolor}
\usepackage{hyperref}
\hypersetup{
	colorlinks = true,
	linkbordercolor={white},
	linkcolor=black,
	citecolor=blue,
	urlcolor=black
}

\DeclareMathAlphabet{\mathcal}{OMS}{cmsy}{m}{n}

\DeclareMathOperator{\sech}{sech}

\newcommand{\xM}{ x - \frac{L_x}{2} }
\newcommand{\yB}{ y- \frac{L_y}{4} }
\newcommand{\yT}{ y- \frac{3L_y}{4} }

\newcommand{\BGzero}{BG0}
\newcommand{\LBG}{BGlow}
\newcommand{\HBG}{BGmed}
\newcommand{\BGthree}{BGhigh}

\newcommand{\yBd}{\dfrac{\yB}{\delta} }
\newcommand{\yTd}{\dfrac{\yT}{\delta} }

\newcommand{\xMdx} { \left( \dfrac{\xM}{ \delta_x}  \right) }
\newcommand{\yBdy} { \left( \dfrac{\yB}{ \delta_y}  \right) }
\newcommand{\yTdy} { \left( \dfrac{\yT}{ \delta_y}  \right) }

\newcommand{\humpX} { \exp\left(- \xMdx^2 \right)}
\newcommand{\humpB} { \exp\left(- \yBdy^2 \right)}
\newcommand{\humpT} { \exp\left(- \yTdy^2 \right)}

\newcommand{\Ppar}{p_{00}} 
\newcommand{\Pperone}{p_{11}}
\newcommand{\Ppertwo}{p_{22}}

\newcommand{\Pparperone}{p_{01}}
\newcommand{\Pparpertwo}{p_{02}}
\newcommand{\Pperonetwo}{p_{12}}

\renewcommand{\vec}[1]{\bm{#1}}

\makeatletter
\def\@email#1#2{%
 \endgroup
 \patchcmd{\titleblock@produce}
  {\frontmatter@RRAPformat}
  {\frontmatter@RRAPformat{\produce@RRAP{*#1\href{mailto:#2}{#2}}}\frontmatter@RRAPformat}
  {}{}
}%
\makeatother

\begin{document}

\preprint{DOI: \href{https://aip.scitation.org/doi/full/10.1063/5.0066397}{10.1063/5.0066397}}

\title[Identification of closure terms using machine learning]{Identification of high order closure terms from fully kinetic simulations using machine learning}
\author{B. Laperre}

\email{brecht.laperre@kuleuven.be}

\author{J. Amaya}%

\author{S. Jamal}

\author{G. Lapenta}
\affiliation{Centre of Mathematical Plasma- and Astrophysics, KU Leuven, Heverlee, 3001, Belgium}%

\begin{abstract}
    Simulations of large-scale plasma systems are typically based on a fluid approximation approach. These models construct a moment-based system of equations that approximate the particle-based physics as a fluid, but as a result lack the small-scale physical processes available to fully kinetic models. Traditionally, empirical closure relations are used to close the moment-based system of equations, which typically approximate the pressure tensor or heat flux. The more accurate the closure relation, the stronger the simulation approaches kinetic-based results. In this paper, new closure terms are constructed using machine learning techniques. Two different machine learning models, a multi-layer perceptron and a gradient boosting regressor, synthesize a local closure relation for the pressure tensor and heat flux vector from fully kinetic simulations of a 2D magnetic reconnection problem. The models are compared to an existing closure relation for the pressure tensor, and the applicability of the models is discussed. The initial results show that the models can capture the diagonal components of the pressure tensor accurately, and show promising results for the heat flux, opening the way for new experiments in multi-scale modeling. We find that the sampling of the points used to train both models play a capital role in their accuracy.
\end{abstract}

\maketitle

\section{Introduction}

Fluid models are the keystone of macroscopic plasma modeling \cite{chapman1990mathematical}. Plasmas at large scales in all areas of application, from nuclear fusion to industrial plasmas and astrophysics, are treated with models based on a small number of low order moments of the velocity distribution: e.g. density, momentum, temperature, pressure. These moment models are derived rigorously from kinetic theory and the macroscopic plasma equations that follow from the integration of the Boltzmann equation \cite{burgers1969flow} or the Vlasov equation. The moments are computed as integrals in the velocity space weighted by a power of the velocity itself, where the density is linked to the zeroth order moment and momentum to the first.

Currently, the challenge lies in computing the closure \cite{levermore1996moment}. The set of equations describing the evolution of the first $N$ moments always depend on the $N+1$th moment (and sometimes higher), but the evolution of the $N+1$th moment is not provided. Traditionally, closure relations express the missing moment(s) as a function of the lower order moments. Many examples of the application of closure exists, and we refer the reader to the literature with examples ranging from simple equations of state (e.g. adiabatic\cite{negulescu2016closure}, isothermal\cite{passot2017electron}) to more complex descriptions of plasmas in strongly coupled, relativistic or quantum degenerate states\cite{haas2010fluid}. 
All closures are typically derived in two ways \cite{braginskii1958transport,helander2005collisional, freidberg2008plasma}. The first approach empirically determines the closure via experiments, similar to how equations describing the ideal gas law or the transport coefficient in fusion devices were discovered. Whereas the second approach relies on theory, particularly theoretical models summarizing kinetic processes, finite Larmor radius effects or Landau damping. The progress in this direction has accompanied the evolution of plasma science, with experiments as prominent as ITER, being based on the latter approach \citep{fasoli2016computational}. 

In this work, we consider the phenomena of reconnection, the process by which the topology of the magnetic field lines changes. Reconnection not only plays a role in the Sun-Earth dynamo, but is a critical phenomena in the solar corona, nuclear fusion and many more types of high-energetic plasma's \citep{yamada1994investigation, yamada2010magnetic}. It has been shown that the kinetic effects play a very important role in accurately depicting phenomena such as collisionless fast reconnection \cite{hesseMagneticReconnectionSpace2020, malakitScalingAsymmetricMagnetic2010, pucciPropertiesTurbulenceReconnection2017}. Currently, the most accurate models for numerical reconnection simulations use the kinetic approach, through particle-in-cell methods. 

However, kinetic simulations are computationally very expensive, making them generally only feasible in small domains \cite{saitoWhistlerTurbulenceParticleincell2008, lapentaElectromagneticEnergyConversion2014}. Models based on the moments of the phase-space distribution of the particles, such as the magnetohydrodynamics (MHD) model, Hall-MHD, and moment models based on the Vlasov equation, are inevitable for large-scale systems such as Earth's magnetosphere, but have to deal with the problem of closure. In the case of collisional plasma, thermal dynamics provides ways to compute accurate closures relations. However, when considering magnetic reconnection in space-plasmas, the large distance between particles brings the problem into the collisionless limit.  

The work of Wang \textit{et al.} (2015)\cite{wangComparisonMultifluidMoment2015} compares a multi-fluid moment model with a particle-in-cell simulations of collisionless magnetic reconnection. They used a 10-moment model, which uses full pressure tensors but needs a closure relation for the heat flux tensor. A local closure for the change in the heat flux tensor was constructed, based on the collisionless global Hammett-Perkins\cite{hammett1990fluid} closure. Even with their simple local, physically fundamental closure, the pressure tensor still evolved accurately when appropriate parameters are selected. Lautenbach and Grauer (2018)\cite{lautenbachMultiphysicsSimulationsCollisionless2018} based themselves on this approach to construct a multi-physical simulation of collisionless plasmas, combining particle-in-cell, Vlasov, 10-moment and 5-moment models to create a night-side simulation of the Earth's magnetosphere.

Because the closure is an unknown equation that has to be learned from observation and simulation, it is a problem that could be tackled with machine learning (ML). Recently, a new line of investigation is emerging: the use of machine learning tools to replace the analytical formulas used so far. The Sparse Identification of Nonlinear Dynamical systems (SINDy)\cite{de2020discovery} method has been used to identify the different terms of the Vlasov equation governing a two-stream instability PIC simulation\citep{alves2020data}. Discovering governing equations from deep learning models has also been done in the case of turbulence simulations\cite{heinonen2020turbulence}, finding new equations coupling the particle flux to the gradient of vorticity by examining the connections made by the deep learning model. Further work in turbulence showed that a numerical gradient model can be replaced by a deep learning model for 2D simulations \cite{rosofskyArtificialNeuralNetwork2020}.

Other authors studied the performance of ML models in replicating known closure equations. The work by Ma \textit{et al.}\cite{ma2020machine} was one of the first to do a proof of principle, where a series of different machine learning models were used to learn the one-dimensional Hammett-Perkins closure for the heat flux in the electrostatic collisionless limit. It showed that ML models could be a viable alternative to the difficult numerical schemes that would otherwise be necessary. This work was extended by Wang \textit{et al.} (2020)\cite{wangDeepLearningSurrogate2020} to a kinetic Landau-fluid closure with collisions. Their ML model was able to learn the closure function, and was subsequently implemented into an existing numerical fluid simulation code, with positive results. The work of Maulik \textit{et al.}\cite{maulikNeuralNetworkRepresentability2020} extends the work of Ma \textit{et al.} to different closure equations relevant in fusion applications. However, these works only made use of data generated from simplified analytical equations of both the heat flux and temperature, and not data retrieved from full simulations or observations. 

Our work aims to take the first steps toward creating a new machine learning model that has learned a local closure relation from simulation data, instead of learning an analytical equation. We proceed by first conducting highly resolved massively parallel kinetic simulations where all processes are represented as accurately as possible. From these simulations, the necessary features are extracted and prepared for the ML models. In total, three models are trained. Two ML models and a baseline model, for which we used a linear regressor. 

The paper is structured as follows. Section \ref{chapter:fluidmodelsandmoments} discusses different existing fluid models and their closure relations, together with the assumptions we make to learn a local closure relation with the machine learning models. Section \ref{chapter:data} discusses the kinetic simulations and how this data is extracted and prepared for the ML models. Section \ref{chapter:models_methods} discusses the different models used in the experiment and how they are evaluated. Section \ref{chapter:results} shows the result of the training and compares it to an existing closure relation. Finally, section \ref{chapter:discussion} discusses the results, together with the applicability of the model.

\section{Fluid models and closure relations}
\label{chapter:fluidmodelsandmoments}

In this section, the most common fluid-based approaches used to simulate magnetic reconnection are discussed, together with the closures used in those models. First the ideal 5-moment model is derived from the Vlasov equation following the work of Wang \textit{et al.}\cite{wangComparisonMultifluidMoment2015}, and different closures are discussed that accompany these under different conditions. Next, the 10-moment moment model is discussed, together with the local closure for the heat flux. Afterwards, the use of machine learning models is motivated, together with the assumptions that were considered during the experiments.

\subsection{Moment model of the Vlasov equation}

Take a collisionless, multi-fluid plasma. Each species $s$ of the plasma is described by a particle distribution function $f_s(t, \vec{x}, \vec{v})$ in phase space, with its evolution in time described by the Vlasov equation. Using Einstein notation, the Vlasov equation can be written as:
\begin{equation}
    \frac{\partial f_s}{\partial t} + v_i \frac{\partial f_s}{\partial x_i} + \frac{q_s}{m_s}\left(E_i + \epsilon_{abi} v_a B_b\right) \frac{\partial f_s}{\partial v_i} = 0,
    \label{eq:Vlasov}
\end{equation}
with $\vec{E}$ is the electric field, $\vec{B}$ the magnetic field, and $\dfrac{q_s}{m_s}$ the charge over mass ratio of the particle species $s$.  $\epsilon_{ijk}$ is the anti-symmetric Levi-Civita symbol, which is equal to $\pm 1$ for an even\slash odd permutation of its indices (1,2,3), and zero otherwise. The evolution of the magnetic and electric fields are governed by the Maxwell equations.

Computing the solution of the Vlasov equation for a particle distribution is computationally very expensive, and instead a fluid approach is used. A moment model is constructed from the Vlasov equation as follows. The $m$-th moment of the particle species $s$ is defined by multiplying the phase-space distribution function $f_s(t, \vec{x}, \vec{v})$  by a basis function $\phi_m(\vec{v})$, $m=0, \dots, M$ and integrating over the velocity $\vec{v}$. Typically for the kinetic equation, the monomial basis functions $\phi_m(\vec{v}) \equiv \vec{v}^m$ is chosen, representing a tensor product. The first four moments are defined as
\begin{align}
    n_s & \equiv \int f_s(t, \vec{x}, \vec{v}) \phi_0(\vec{v}) d\vec{v} = \int f_s(t, \vec{x}, \vec{v}) d\vec{v} \label{eq:othordermoment} \\
    n_s \vec{u}_s &\equiv \int f_s(t, \vec{x}, \vec{v}) \phi_1(\vec{v}) d\vec{v} = \int f_s(t, \vec{x}, \vec{v}) \vec{v} d\vec{v} \label{eq:1stordermoment} \\
    \frac{1}{m_s}\mathcal{P}_s & \equiv \int f_s(t, \vec{x}, \vec{v}) \phi_2(\vec{v}) d\vec{v} = \int f_s(t, \vec{x}, \vec{v}) \vec{v}^2 d\vec{v} \\
    \frac{1}{m_s}\mathcal{Q}_s & \equiv \int f_s(t, \vec{x}, \vec{v}) \phi_3(\vec{v}) d\vec{v} = \int f_s(t, \vec{x}, \vec{v}) \vec{v}^3 d\vec{v} \label{eq:4thordermoment}
\end{align}
with $n_s(t, \vec{x})$ the number density and $\vec{u}_s(t, \vec{x})$ the mean velocity. The second and third order moment are the energy density tensor and the heat flux tensor, respectively. The evolution of these moments is extracted from the Vlasov equation as follows: by multiplying Eq. \eqref{eq:Vlasov} with the monomial test functions $\varphi_m(\vec{v}) \equiv v^m, m=0,\dots,M$ and integrating over $\vec{v}$, the exact moment equations for the $m$-th order moment are found for each species. The first three evolution equations are (dropping the $s$ for brevity):
\begin{align}
    \frac{\partial n}{\partial t} + \frac{\partial}{\partial x_j}(nu_j) &= 0 \label{eq:1moment}\\
    m\frac{\partial (n u_i)}{\partial t}    + \frac{\partial \mathcal{P}_{ij} }{\partial x_j} &= q (nE_i + n \epsilon_{ijk} u_j B_k) \label{eq:2dmoment}\\
    \frac{\partial \mathcal{P}_{ij}}{\partial t}   + \frac{\partial \mathcal{Q}_{ijk}}{\partial x_k} &= 2qu_{(i} E_{j)} + n \epsilon_{(ikl} \mathcal{P}_{kj)} B_l \label{eq:3dmoment}
\end{align}
The brackets in the indices represent symmetric contraction of the indices, so for example $u_{(i} E_{j)} = u_i E_j+u_j E_i$. Finally, define the heat flux and the pressure tensor from $\mathcal{P}$ and $\mathcal{Q}$, respectively:
\begin{align}
    Q_{ijk} &\equiv m \int (v_i-u_i)(v_j-u_j)(v_k-u_k)f(t, \vec{x}, \vec{v}) d\vec{v} = \mathcal{Q}_{ijk} - 2nmu_iu_ju_k - u_{(i} \mathcal{P}_{jk)} \label{eq:heatfluxtensor} \\
    P_{ij} &\equiv m \int (v_i-u_i)(v_j-u_j)f(t, \vec{x}, \vec{v}) d\vec{v} = \mathcal{P}_{ij} - nmu_iu_j \label{eq:pressuretensor}.
\end{align}
From Eqs. \eqref{eq:1moment}-\eqref{eq:3dmoment}, together with the Maxwell equations, different fluid models can be constructed. Let us start with the ideal 5-moment model. Assuming isotropic pressure, define the scalar pressure as $p \equiv \frac{1}{3}P_{ii}$ and the total fluid energy as the trace of the energy density tensor $\mathcal{P}$:
\begin{equation}
    \mathcal{E} \equiv \frac{1}{2}\mathcal{P}_{ii} = \frac{3}{2} p + \frac{1}{2} mn\vec{u}^2
\end{equation}
Taking the trace from Eq. \eqref{eq:3dmoment} then gives us the evolution of the energy in time:
\begin{equation}
    \frac{\partial \mathcal{E}}{\partial t} + \frac{1}{2}\frac{\partial \mathcal{Q}_{iik}}{\partial x_k} =  nq\vec{u} \cdot \vec{E} .
\end{equation}
What remains now is the heat flux tensor component. From Eq. \eqref{eq:heatfluxtensor} it follows that 
\begin{equation}
    \frac{1}{2} \mathcal{Q}_{iik} = q_k + u_k (p + \mathcal{E}) + u_i \pi_{ik}
\end{equation}
where $q_k \equiv Q_{iik}/2$ is the heat flux vector and $\pi_{ij} \equiv P_{ij}-p\delta_{ij}$ the viscous stress tensor. The ideal five moment model is obtained by using the closure $\pi_{ij} = q_k = 0$. The evolution of the five moments are given by Eqs. \eqref{eq:1moment}, \eqref{eq:2dmoment} and 
\begin{equation}
    \frac{\partial \mathcal{E}}{\partial t} + \frac{\partial}{\partial x_k}\left( u_k (p + \mathcal{E}) \right) = nq\vec{u} \cdot \vec{E}.\label{eq:energyevolution5moment} 
\end{equation}
The 5-moment model is more general than the MHD model or Hall MHD model\cite{hakimTwofluidPhysicsFieldreversed2007, wangComparisonMultifluidMoment2015}. It takes into account electron inertia, charge separation and the full electromagnetic field equations, together with separate electron and ion motion and (adiabatic) pressure. Wang \textit{et al.}\cite{wangComparisonMultifluidMoment2015} showed that the 5-moment model, in the limit of vanishing electron inertia, infinite speed of light and quasi-neutrality, approaches the Hall MHD equations. 

If the plasma is not isotropic, which is the case for the magnetotail reconnection, as shown by by Egedal \textit{et al.}\cite{egedal2008evidence}, the pressure tensor can no longer be approximated as the scalar pressure. Le \textit{et al.}\cite{le2009equations} showed that for collisionless guide-field reconnection, the following approximation for $p_\perp$ and $p_\parallel$ can be used:
\begin{align}
    \tilde{p}_{*\parallel} &= n_* \frac{2}{2+\alpha} + \frac{\pi^3 n_*^3}{6B^2_*}\frac{2\alpha}{2\alpha + 1} \label{eq:parallel_p_closure}\\
    \tilde{p}_{*\perp} &= n_* \frac{1}{1+\alpha} + n_* B_* \frac{\alpha}{\alpha + 1} \label{eq:perp_p_closure}
\end{align}
where $\alpha = n^3_*/B^2_*$, and for any quantity $X$ define $X_* = X/X_\infty$, where $X_\infty$ is the value of $X$ upstream of the reconnection region. The pressure tensor itself is then approximated with 
\begin{equation}
    P_{ij} = p_\perp \delta_{ij} + (p_\parallel - p_\perp) \frac{B_i B_j}{B_{ii}^2}.
\end{equation}
This closure approaches the double-adiabatic forms for high density, and used by Ohia \textit{et al.}\cite{ohiaDemonstrationAnisotropicFluid2012} in combination with a two-fluid MHD formulation to simulate magnetic reconnection. 

Finally, the 10-order moment model starts from Eqs. \eqref{eq:1moment} - \eqref{eq:3dmoment} The full pressure tensor $P$ is evolved, but a closure relation for the divergence of the heat flux tensor is necessary to close this set of equations. Hammett and Perkins\cite{hammett1990fluid} showed that by using a linear approximation, the perturbation of the heat flux in 1 dimension can be related to the temperature in Fourier space:
\begin{equation}
    \tilde{q}(k) = -n_0 \chi_1 \frac{\sqrt{2}}{|k|} ikv_t \tilde{T}(k), \label{eq:HammettPerkinsClosure}
\end{equation}
with tildes indicating perturbations around equilibrium, $k$ the wave number, $v_t = \sqrt{T/m}$ the thermal velocity, $n_0$ the number density at equilibrium and $\chi_1$ some constant. Taking the Fourier inverse returns an integral of the temperature along the field lines, making this closure non-local. 

\subsection{Machine learning and closure}
\label{sect:MLandClosure}
Machine learning models learn relations between a set of inputs and outputs by changing their internal parameters, based on the error between the expected output and the model's output. The most common model is the multi-layer perceptron (MLP), and has been applied in many fields of physics already\cite{camporealeChallengeMachineLearning2019, carleo2019machine}. For details on the different concepts and models that exist, the interested reader is referred to the excellent book by Goodfellow\cite{goodfellow2016deep}.

The goal of this work is to provide the first steps toward learning a non-local closure to the full heat flux tensor from simulation data. The work by Ma \textit{et al.}\cite{ma2020machine} showed that an ML model can learn the non-local Hammett-Perkins closure, but this was done in the 1-dimensional case, with data constructed from simplified noise-free analytical equation. In this work, we show that an ML model can learn a local closure approximation for the electrons pressure tensor $P_e$ and the electrons heat flux vector $q_e$. The pressure tensor computed by the ML model will be compared to the empirical closure of the pressure tensor shown in Eqs.~\eqref{eq:parallel_p_closure}-\eqref{eq:perp_p_closure}.

We assume that the pressure tensor and heat flux vector can both be learned from local features. While the closure described by equations \eqref{eq:parallel_p_closure}-\eqref{eq:perp_p_closure} has some non-locality due to the ratio with the values upstream of the reconnection region, it can still largely be seen as a local closure. Only the closure for the heat flux, described by Eq.~\eqref{eq:HammettPerkinsClosure}, is non-local due to the integration over the field lines that is needed for the Fourier transformations. However, Wang \textit{et al.}\cite{wangComparisonMultifluidMoment2015} argues that Eq.~\eqref{eq:HammettPerkinsClosure} can be reduced to a local approximation by replacing the wave number $k$ by a typical wave-number $k_0$, related to the scale on which the collisionless damping occurs. They used this local approximation successfully to simulate collisionless reconnection. In this work, our methods learn the heat flux vector from local features only. Learning a non-local closure with ML, and the complete heat flux tensor, will be examined in a future work. 


\section{Simulation and model Data}
\label{chapter:data}

In this section, first the details of the simulations are discussed. Then the model inputs are given, and finally we describe how the  training, validation and test set are constructed from the simulation data.

\subsection{Simulation data}

The data used to train the models is extracted from kinetic simulations of magnetic reconnection. The simulation, denoted in our work as "Double-Harris sheet experiment", is the same experimental setup described in Markidis \textit{et al.} (2010)\cite{markidis2010multi}. The implicit Particle-in-Cell (PiC) code iPiC3D\cite{ricci2002simplified} is used for the simulations.

The size of the simulation is set to $L_x \times L_y \times L_z =30 d_i \times 40 d_i \times 0.1 d_i$, where $d_i$ is the ion inertial length. Each spatial direction has respectively $769 \times 1025 \times 1$ cells, with periodic boundary conditions. The simulation evolves with a time step equal to the inverse of the ion cyclotron frequency $\Delta t = \omega_{i,c}^{-1} = 0.0625$. 
Four species of particles are used, 2 electron species and 2 ion species. The mass ratio between the ions and the electrons is set to $m_{i}/m_{e}= 256$. For more details on the simulation setup, we refer to Appendix \ref{appendix:simulation}. 
Although the physical mass ratio would give more accurate simulations, this would come at a cost of higher computation time. The current choice in mass-ratio gives a good trade-off between accuracy and simulation time, ideal for the scope of this paper. Similar mass-ratio's were used in different papers to demonstrate closure models\cite{le2009equations, wangComparisonMultifluidMoment2015, hesseRoleElectronHeat2004} 

Four kinetic simulations are created, each with identical initial conditions (as described in Appendix \ref{appendix:simulation}), except for the guiding background magnetic field $B_{0z}$. The four simulations are named based on the choice of their guiding field, and displayed in Table \ref{table:guidingfield}. The different background magnetic field ensures that, while the simulation is initially the same, different physics will arise in and around the region of reconnection. 

\begin{table}[ht]
    \caption{The guiding and background magnetic field of each of the four iPiC3D simulations in dimensionless units.}
	\label{table:guidingfield}
    \begin{ruledtabular}
    	\begin{tabular}{lcccc}
    	{Simulation}	&\textbf{ \BGzero\ }& \textbf{\LBG\ } & \textbf{\HBG\ } & \textbf{\BGthree\ } \\
    	\hline
    	$B_{x0}$ & 0.0097 & 0.0097 & 0.0097 & 0.0097 \\
    	$B_{y0}$ & 0 	  & 0	& 0	 & 0\\
    	$B_{z0}$ & 0 & 0.00097 & 0.0097 & 3 \\
    	\end{tabular}
    \end{ruledtabular}
\end{table}

From each simulation, the moments of the phase-space particle distribution described in section \ref{chapter:fluidmodelsandmoments} are computed using B-splines\cite{knott2000interpolating, birdsall2018plasma, hockney2021computer}. These transfer the information from the particles at $x_p$ to the cells on $x_g$, denoted as $S(\vec{x}_g - \vec{x}_{p})$. For each particle species $s$, located at position $x_p$, the charge density $\rho$, the mean velocity $u$, the pressure tensor $P$ and the heat flux vector $q$ are computed on the uniform grid $x_g$:
\begin{eqnarray}
    \rho_{sg} \equiv \rho_s( \vec{x}_g)  &=& \sum^{N_{sg}}_{p_s \in \Omega_g} q_s S(\vec{x}_g - \vec{x}_{p}), \\
    \vec{u}_{sg} \equiv \vec{u}_s(\vec{x}_g, t) &=& \frac{1}{\rho_{sg}} \sum^{N_{sg}}_{p_s \in \Omega_g} \vec{v}_{p} S(\vec{x}_g - \vec{x}_{p}) , \\
    P_{sg} &=& \sum^{N_{sg}}_{p_s \in \Omega_g}   m_p (\vec{v}_p - \vec{u}_p)  (\vec{v}_p - \vec{u}_p)  S(\vec{x}_g-\vec{x}_{p}), \\
    \vec{q}_{sg} &=& \sum^{N_{sg}}_{p_s \in \Omega_g} m_p (\vec{v}_p - \vec{u}_p)^2 (\vec{v}_p - \vec{u}_p)  S(\vec{x}_g-\vec{x}_{p}).
\end{eqnarray}
Here, $\vec{x}_p$, $\vec{v}_p$ and charge $q_p$ are respectively the position, velocity and charge of the particle $p$ of species $s$. 
The summation is computed over $N_s$, the total number of particles of species $s$, with $\Omega_g$ the collocated cell to which $p_s$ belongs. 
The magnetic and electric field are also extracted from the simulation. Although all the quantities are physical, the quantities extracted from the simulation have all been transformed to dimensionless code units. The code unit transformation is described in Appendix \ref{appendix:codeunits}. Throughout the paper, the normalized physical quantities are used.

The pressure tensor $P$ is transformed to the field-aligned basis~\cite{goldmanWhatCanWe2016}. 
Define the new unit vectors of this basis as
\begin{equation}
\hat{\vec{e}}_\parallel = \frac{\vec{B}}{\|\vec{B}\|}, \quad \hat{\vec{e}}_{\perp,1} = \frac{\vec{B} \times \hat{\vec{z}}}{\|\vec{B} \times \hat{\vec{z}}\|}, \quad 
\hat{\vec{e}}_{\perp, 2} = \hat{\vec{e}}_b \times\hat{\vec{e}}_{\perp,1}.
\end{equation}
Let $\{\hat{\vec{e}}_\parallel, \hat{\vec{e}}_{\perp,1}, \hat{\vec{e}}_{\perp, 2}\}$ be column vectors. Then the field-aligned pressure tensor is defined as
\begin{equation}
\label{eq:field_aligned_pressure}
P^B_{sg} =  \begin{pmatrix} \hat{\vec{e}}_\parallel & \hat{\vec{e}}_{\perp,1} & \hat{\vec{e}}_{\perp, 2} \end{pmatrix} P_{sg} \begin{pmatrix} \hat{\vec{e}}_\parallel & \hat{\vec{e}}_{\perp,1} & \hat{\vec{e}}_{\perp, 2} \end{pmatrix}^T.
\end{equation}
We name the diagonal components of $P$ as $\Ppar, \Pperone$ and $\Ppertwo$ and the off-diagonal components are given the names $\Pparperone, \Pparpertwo, \Pperonetwo$, defined as
\begin{align}
\text{Diagonal:} \quad \begin{cases}
\Ppar = \hat{\vec{e}}_\parallel \  P_{sg} \ \hat{\vec{e}}_\parallel^T \\
\Pperone = \hat{\vec{e}}_{\perp,1} \  P_{sg} \   \hat{\vec{e}}_{\perp,1},  \\
\Ppertwo = \hat{\vec{e}}_{\perp,2} \   P_{sg} \  \hat{\vec{e}}_{\perp,2}
\end{cases}  & \text{ Off-diagonal:} \quad
\begin{cases}
\Pparperone = \hat{\vec{e}}_\parallel \  P_{sg} \ \hat{\vec{e}}_{\perp,1}^T \\
\Pparpertwo = \hat{\vec{e}}_\parallel \  P_{sg} \   \hat{\vec{e}}_{\perp,2}  \\
\Pperonetwo = \hat{\vec{e}}_{\perp,1} \   P_{sg} \  \hat{\vec{e}}_{\perp,2}
\end{cases}.
\end{align}
Note that there are only three off-diagonal components, as the pressure tensor is symmetric.

\subsection{Model input and output}
\label{sect:data:input}
In this section, the choice of physical quantities as input, and the physical quantities we which to predict, are given. The targets of the models are the three components of the heat flux vector $q$ and the six components that form the electron pressure tensor $P_e$:
\begin{itemize}
    \item The three heat flux vector components $\vec{q} = \{q_x, q_y, q_z\}$
    \item The three diagonal components of $P_e$: $\Ppar, \Pperone, \Ppertwo$.
    \item The three off-diagonal components of $P_e$: $\Pparperone, \Pparpertwo, \Pperonetwo$.
\end{itemize}

The input of the ML models are a vector consisting of the magnetic field $B$ together with its gradients and magnitude, the mean velocity $u$, the charge density $\rho$ and the value $\alpha$, based on the closure from Le \textit{et al.}\cite{le2009equations}. The value $\alpha$ is included explicitly due to its correlation to the perpendicular pressure $p_\perp$\cite{le2009equations}. The gradients of $B$ are computed using a second order finite difference. Finally, the input features that are highly linearly correlated to other input features are removed from the input, in order to speed up the training process and make the models more interpretable. Following these steps, 14 input features are chosen:
\begin{itemize}
	\item The magnetic field $\vec{B}$ (3), $||\vec{B}||$ (1), and its gradients, with the exception of $\partial_x B_y$ (5)
	\item The velocity vector $\vec{u}$ (3)
	\item The density $\rho_e$ (1)
	\item The value $\alpha = \frac{\rho_e^3}{||\vec{B}||^2}$ (1), taken from the closure of Le \textit{et al.}\cite{le2009equations}, see Eqs. \eqref{eq:parallel_p_closure}-\eqref{eq:perp_p_closure}.
\end{itemize}

\subsection{Training, validation and test set preparation}
\label{sect:create_datasets}
\begin{figure}[ht]
	\centering
	\includegraphics[]{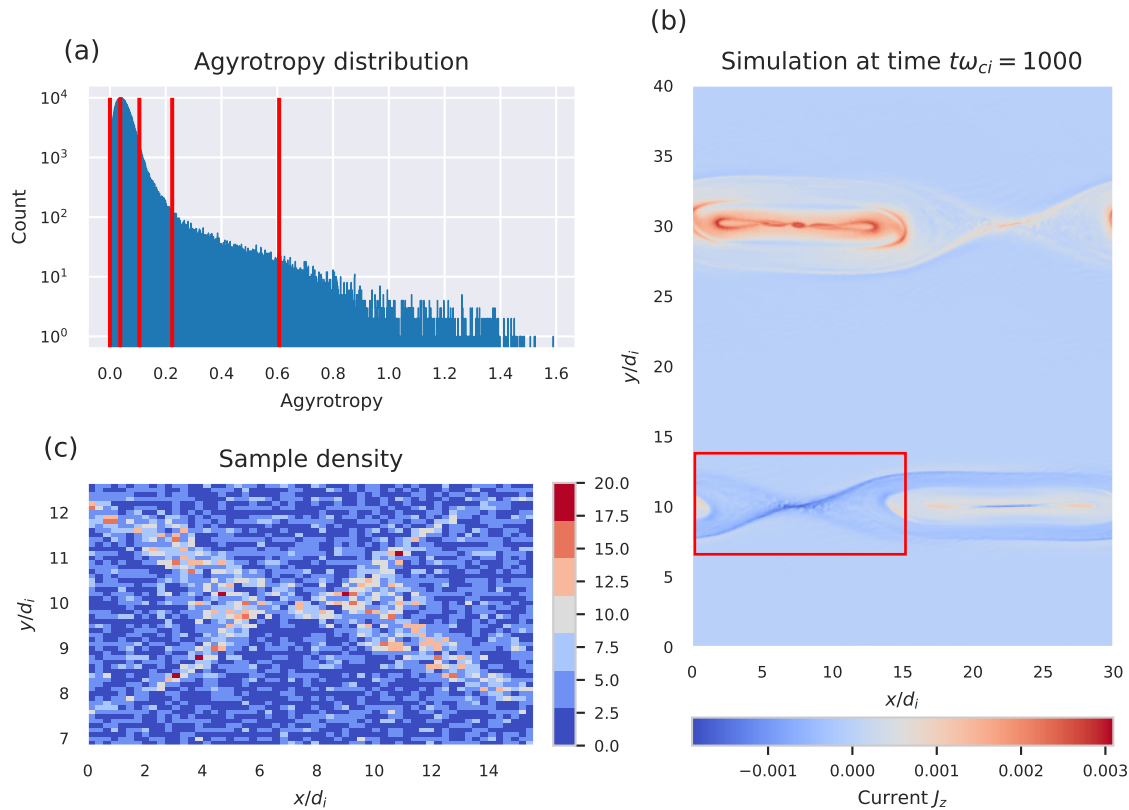}
	\caption{(a): The distribution of the values of the agyrotropy from a single snapshot of the simulation. High agyrotropy corresponds to reconnection regions. Almost all of the cells in the snapshot have an agyrotropy between 0 and 0.2. The red lines indicate the bin boundaries. The size of each bin increases exponentially, to account for the exponential decay in the number of cells with a high agyrotropy. (b): The full simulation. The region circled in red is the reconnection region from which the data points are sampled.
	(c): An example of the cells sampled from the distribution shown in (a) at the reconnection site highlighted in (b). A higher sampling density is seen at the reconnection site, because of their high agyrotropy value.}
	\label{fig:agyro_dist}
\end{figure}

The training, validation and test set are constructed by sampling cells from snapshots of the four simulations around the reconnection region. In order to sample a balanced number of cells from regions with and without reconnection, the agyrotropy of the electrons is used. The agyrotropy is computed in each cell from the eigenvalues of the field-aligned pressure tensor\cite{scudder2008illuminating}. Let  $\lambda_i, \  i \in \{1,2,3\}$ be the eigenvalues of the field-aligned electron pressure tensor $P^B_{eg}$, with $\lambda_1 > \lambda_2 > \lambda_3$. Then the agyrotropy $Ag$ is computed as 
\begin{equation}
Ag = 2\frac{|\lambda_3 - \lambda_2|}{\lambda_3 + \lambda_2}. \label{eq:agyrotropy}
\end{equation}
When the pressure tensor is gyrotropic, the agyrotropy is close to 0. An agyrotropic pressure tensor (which occurs at regions close to or at reconnection) will have a larger agyrotropy. A histogram of the agyrotropy for one of the snapshots of the simulations can be seen in Fig. \ref{fig:agyro_dist}a, and shows that most cells have an agyrotropy close to zero. By dividing the agyrotropy into a set of intervals called bins, as displayed by the red lines in Fig. \ref{fig:agyro_dist}a, we prevent the reconnection regions from being undersampled compared to the regions outside of the reconnection zone. Fig. \ref{fig:agyro_dist}c shows the sampling density after sampling an equal number of cells from each bin. When the sampling is performed, the sampling algorithm is allowed to take duplicates of cells. This is necessary for creating a balanced data set. Only the test set has no duplicates.
The full process of how the training, validation and test sets are extracted and created from the data is described in detail in Appendix \ref{appendix:datasetconstruction}. A training, validation and test set is created for each of the four simulations, and respectively contain \num{16000} cells, \num{4500} cells and \num{7500} cells.

\section{Models and methods}
\label{chapter:models_methods}

\begin{figure*}
    \centering
    \includegraphics[width=0.9\textwidth]{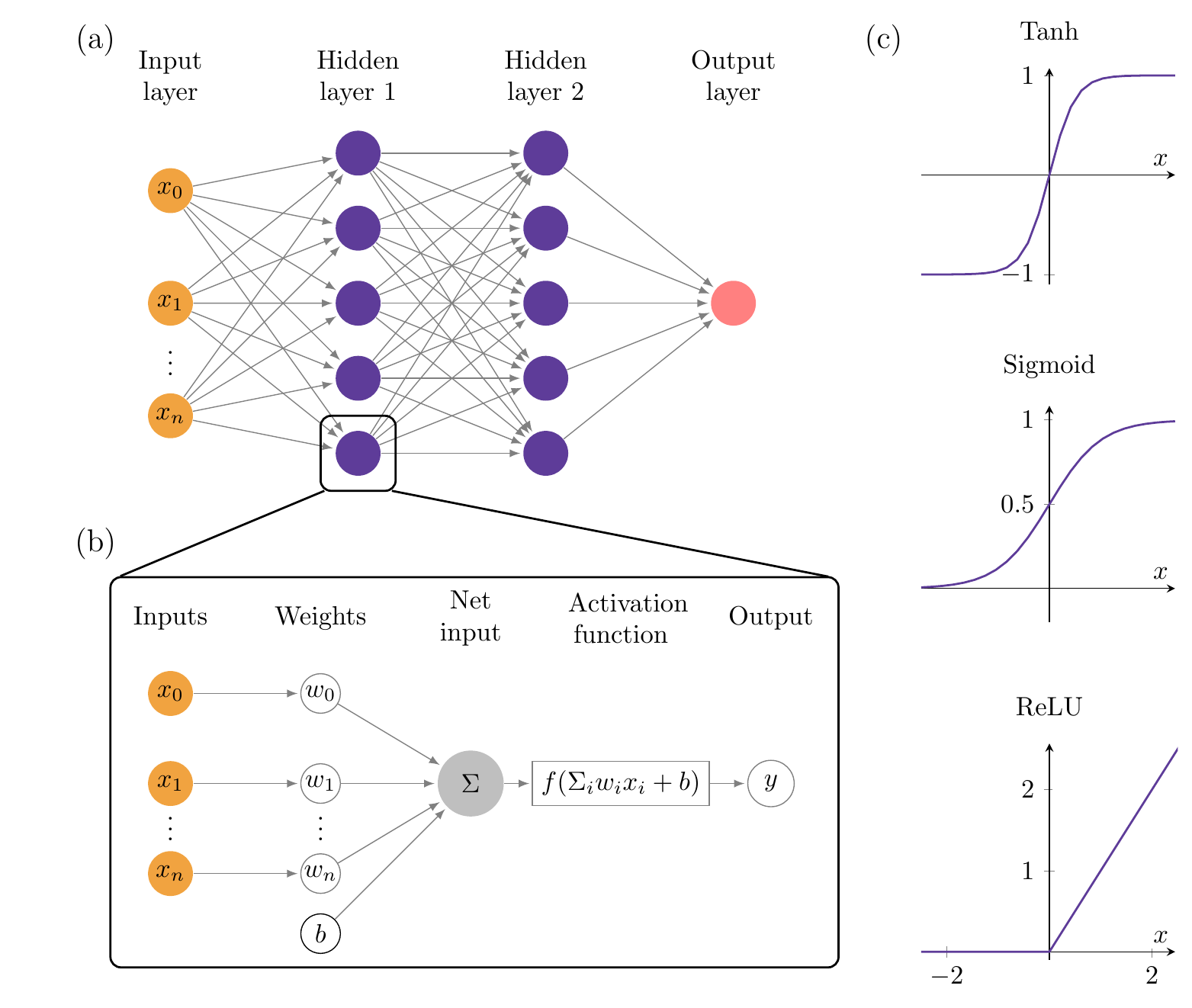}
    \caption{(a): The layout of a standard MLP with two hidden layers. (b): The layout of a single perceptron. The perceptron accepts a set of inputs, multiplies it with its internal weight vector and runs it through its activation function. It returns the scalar output $y$. (c): Examples of activation functions used in perceptrons. }
    \label{fig:MLP_layout}
\end{figure*}

The first machine learning model is a feed-forward multi-layer perceptron (MLP). This type of model has also been used in many related works\cite{ma2020machine, wangDeepLearningSurrogate2020, heinonen2020turbulence, rosofskyArtificialNeuralNetwork2020}. As the name implies, the MLP consists of multiple interconnected layers of a smaller, simpler model, the perceptron, as shown in Fig. \ref{fig:MLP_layout}a. Each perceptron is a simple model that takes an input vector $\vec{x}$, multiplies it with the perceptron's internal weight vector, and runs it through an activation function. This process is also shown in Fig. \ref{fig:MLP_layout}b. The activation function is typically a threshold-like function, such as a $\tanh$, a logistic function or a Rectified Linear Unit (ReLU).
An MLP can have many configurations, and the number of layers, the number of perceptrons per layer, the perceptrons activation function, etc., are all a free choice for the user. The exact configuration of the model used in the experiments can be found in Appendix \ref{appendix:hyperparameters}.
The model was trained using the Adam\cite{kingma2014adam} algorithm for 400 epochs. The validation set is used to prevent over-fitting, a method called 'early-stopping'. Over-fitting is the process where the model starts memorizing the training set instead of learning the dynamics behind the training set. The interested reader is referred to the excellent book by Goodfellow \textit{et al.}\cite{goodfellow2016deep} for more details and information on machine learning models. The Python package PyTorch\cite{paszke2019pytorch} is used to implement the MLP model. 

The second machine learning model is a Gradient Boosting algorithm\cite{friedman2001greedy}. This type of model is based on the principle of ensemble learning\cite{breiman2001random, dietterich2002ensemble}, which combines many weak learners to form a strong learner. Our model is constructed out of many decision trees\cite{quinlan1986induction}. In this paper, the Histogram Gradient Boosting Regressor\cite{ke2017lightgbm} (HGBR) model implemented in the Scikit-Learn\cite{pedregosa2011scikit} package is used for the experiments.

The third and final model is a baseline model, for which a linear regressor is used. The idea of a baseline is based on Occam's Razor, stating that given two plausible options, the simpler option is most often the correct one. The linear regression model is used to assess if a complex ML model is a better performing model. If the performance of the linear regressor is comparable to that of the ML model, we can conclude that the linear regressor is the preferable model, since it is simpler and easier to understand. The linear regression method implemented in the Python package Scikit-Learn is used. 

The code and experiments will be made available after acceptance on Github, at \url{https://github.com/brechtlaperre}.

\subsection{Hyperparameter tuning}

As stated previously, machine learning models have a large set of free parameters, called hyperparameters, that determine both their architecture and method of training. For example, the number of hidden layers (HL) and the number of perceptrons per layer of an MLP. Finding the optimal hyperparameters is an important and problem-dependent problem. The right choice of hyperparameters can have large effects on both accuracy and convergence of the model. There is no formula that provides the right hyperparameters for a given problem, so these must be searched by trial and error. This procedure can be optimized using an iterative Bayesian search over a range of models with different hyperparameters. Table \ref{tab:hyperparams} provides a list of the hyperparameters that were optimized, for both the MLP and the HGBR model, together with the range in which the optimal value was searched for. The Python package Optuna\cite{akiba2019optuna} was used to perform this search.

\begin{table}[ht]
    \centering
    \caption{Hyperparameters of the two machine learning models. Only the training epochs of the MLP model was not determined by the hyperparameter search. The chosen hyperparameters can be found in Appendix \ref{appendix:hyperparameters}.}
    \label{tab:hyperparams}
    \begin{ruledtabular}
	\begin{tabular}{ll|ll}
       MLP & Range & HGBR & Range \\
       \hline
       Training epochs & 200 (Fixed)  & Max iteration & [100 - 300] \\
       Batch size & [120-1024] & Max tree depth & [2 - 20] \\
       Learning rate & [1 - $10^{-6}$] & Learning rate & [1 - $10^{-3}$] \\
       Number of HL & [5 - 7] & Loss function & [least square, least absolute deviation] \\
       HL size & [150 - 600] & & \\
       HL activation function & [Tanh, ReLU] & & \\
       Dropout probability & [0.2 - 0.6] & & 
    \end{tabular}
    \end{ruledtabular}
\end{table}

\subsection{\label{sect:metrics} Evaluation metrics}

A set of metrics is defined to evaluate the models. The evaluation of the models happens on the native scale of the output features, so any transformations applied during the construction of training and test sets are reversed before evaluation. Let $M_i$ be the value predicted by the model at cell $i$, and $O_i$ the observed (true) value at cell $i$, with $i = 1, \dots, N_c$, and $N_c$ the total number of cells in the simulation or data set. 

To quantify the model's ability to predict the variations in the output features, we define the prediction efficiency, also called the $R^2$-score:
\begin{equation}
\label{eq:eval:PE}
R^2 = 1 - \frac{\sum_{i=1}^{N_c} (M_i - O_i)^2 }{\sum_{i=1}^{N_c} (O_i - \bar{O})^2 },
\end{equation}
where $\bar{O} = \frac{1}{N_c}\sum_{i=1}^{N_c} O_i$ is the average of the observed values. The range of the prediction efficiency is $R^2 \in ]-\infty, 1]$, where $R^2=1$ corresponds to a perfect prediction by the model\citep{liemohn_rmse_2021}. 

Next, the relative error of a point $i$ in percentage is defined as
\begin{equation}
    \text{Rel. error} = \frac{| M_i - O_i |}{O_i} \cdot 100
    \label{eq:rel_error}
\end{equation}

Because the model is learning a closure relation which is, by definition, an approximation, we expected the models to never have $R^2=1$. In such a case, it is recommended to check the model on overfitting. However, the closer it is to the most accurate value, the better we expect the model to work.

\section{Experiment and results}
\label{chapter:results}
First, the models are trained on the training sets constructed from the four simulations, as described in Section \ref{sect:create_datasets}. The models are evaluated on both their performance on the training and the test set with the $R^2$ metric. Then, the models are evaluated on a complete reconnection region. 
Finally, the ML models are evaluated along slices in the reconnection region and compared against the closure model by Le et al, defined by Eqs. \eqref{eq:parallel_p_closure} - \eqref{eq:perp_p_closure}.

\subsection{Comparing model performances}
\label{section:comparingPerformances}
\begin{figure}
    \centering
    \includegraphics[width=\textwidth]{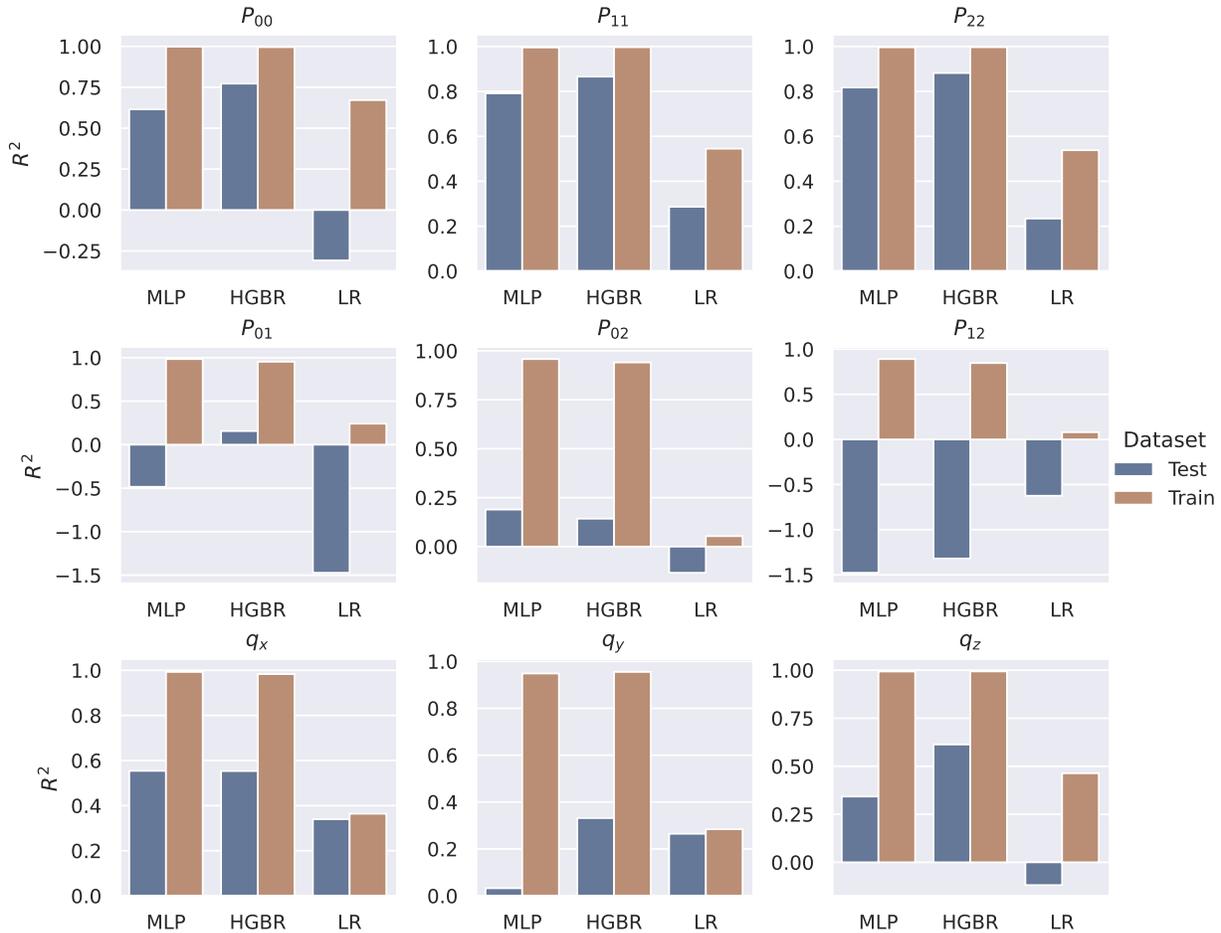}
    \caption{$R^2$ value for each model on both the training and test set. The first row depicts the performance on the diagonal components of the pressure tensor, the second row on the off-diagonal components of the pressure tensor, and the final row on the heat flux vector.}
    \label{fig:R2-comparison}
\end{figure}

The MLP, HGBR and linear regressor are trained on the combined training sets from each simulation. The $R^2$ of each model can be found in Fig. \ref{fig:R2-comparison}. First the performance of the MLP and HGBR model is discussed. A high $R^2$ is seen on both the MLP and HGBR model on the predictions of the diagonal components of the pressure tensor. 
The performance on the test set for the off-diagonal components have lower $R^2$ values, especially compared to the strong performance on the training set. Although the $R^2$ value of the linear regressor is close to zero on the training set, the performance on the test set is the best for the $P_{12}$ component. Finally, the performance on the heat flux vector of both ML models is better than the linear regressor, but the performance is not as strong when compared to the diagonal components of the pressure tensor. Finally, the HGBR model has a consistently higher $R^2$ performance on the test set compared to the MLP model. 

The linear regressor has the weakest $R^2$ performance, and the ML models always perform better on the training set. This behavior is repeated for the test set, and only for two exceptions does the linear regressor perform better. For the off-diagonal component $P_{12}$ and the heat flux vector component $q_y$, the linear regressor shows a higher $R^2$ value on the test set compared to the ML models. A possible explanation is that the ML models are overfitting on the training set for these two components, which results in a performance on the test set that is even weaker than the linear regressor.  

\begin{figure}
    \centering
    \includegraphics[width=\textwidth]{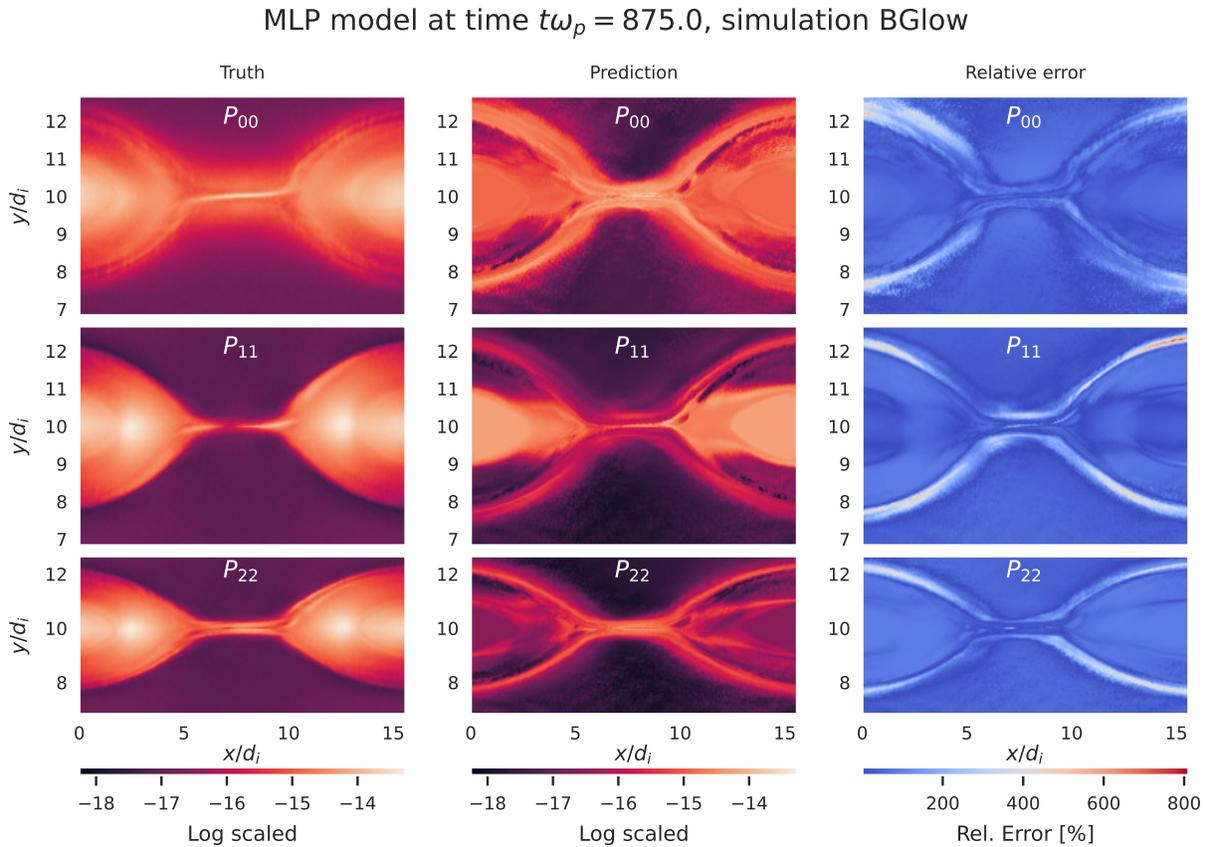}
    \caption{Prediction of the MLP model for the diagonal components of the pressure tensor. The first column shows the log-scaled correct values, the second column the log-scaled prediction by the MLP model, and the third model shows the relative error as defined by Eq. \eqref{eq:rel_error}.}
    \label{fig:MLP_truth_pred_err}
\end{figure}

\begin{figure}
    \centering
    \includegraphics[width=\textwidth]{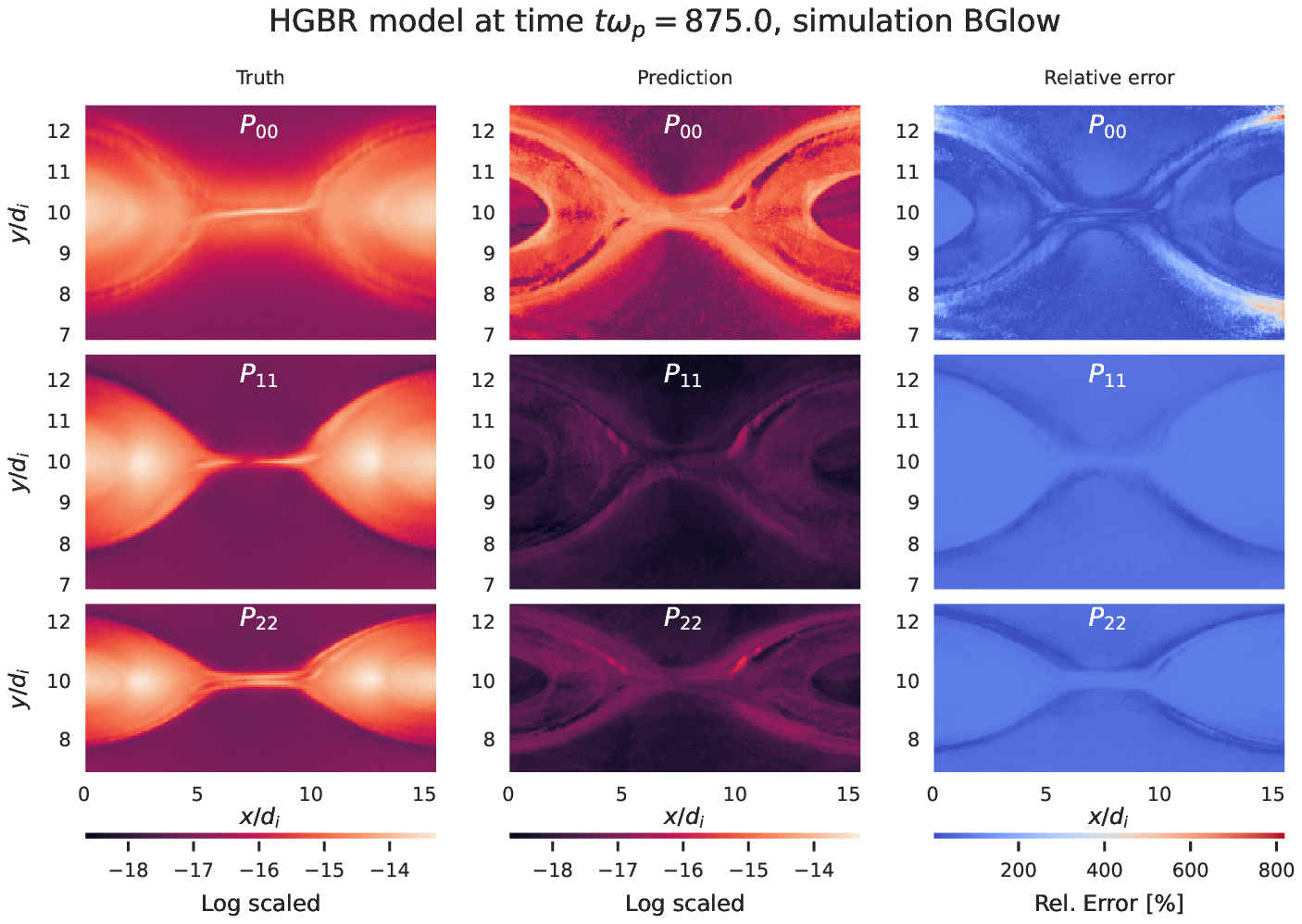}
    \caption{Prediction of the HGBR model for the diagonal components of the pressure tensor. The first column shows the log-scaled correct values, the second column the log-scaled prediction by the HGBR model, and the third model shows the relative error as defined by Eq. \eqref{eq:rel_error}.}
    \label{fig:HGBR_truth_pred_err}
\end{figure}

Finally, the performance of the models are evaluated on a full reconnection region. The snapshot of the simulation at time $t\omega_p = 875$ is taken, and the region spanning $x \in [0, 15], y \in [7, 12]$ is considered. In every simulation, this region develops a reconnection region. First, the relative error between the model prediction and the simulation results are observed. For this purpose, the results on the diagonal components of the pressure tensor are visualized. This can be seen in Fig. \ref{fig:MLP_truth_pred_err} for the MLP model and in Fig. \ref{fig:HGBR_truth_pred_err} for the HGBR model. 

The results for the MLP in Fig. \ref{fig:MLP_truth_pred_err} show that the model can recreate the X-lines and the  structure of the reconnection zone. The relative error shows that the largest error is located in the boundary between the reconnection region and the region outside reconnection. 

The results of the HGBR model in Fig. \ref{fig:HGBR_truth_pred_err} show that the reconnection structure is retrieved for the $\Ppar$ direction of the pressure tensor. However, the prediction of the remaining two components show that the HGBR is underestimating the values of pressure tensor. While the reconnection structure is still visible, the values are close to zero. 

\begin{figure}
    \centering
    \includegraphics[width=\textwidth]{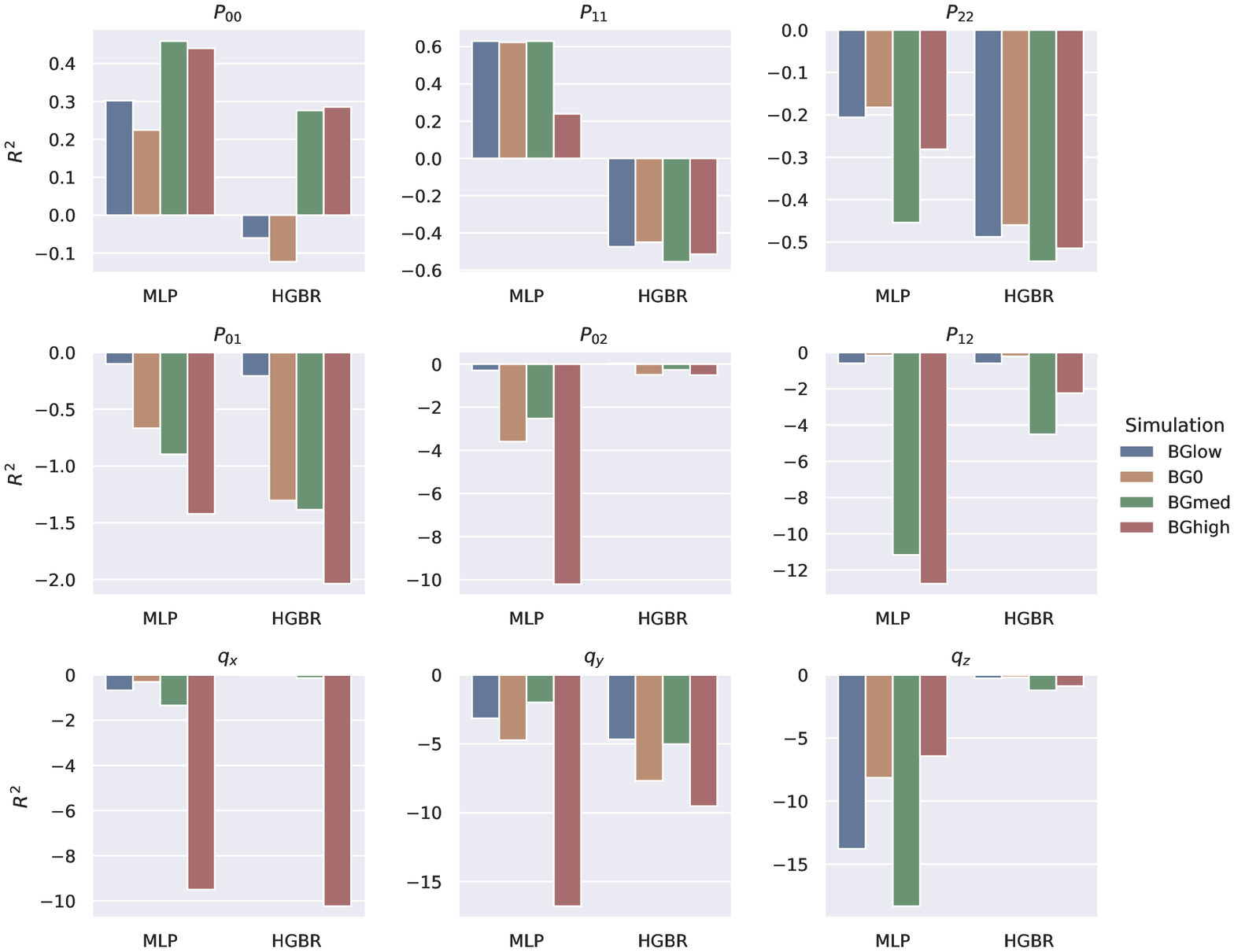}
    \caption{The $R^2$ performance of each model's prediction on the reconnection region displayed in Fig. \ref{fig:MLP_truth_pred_err} and \ref{fig:HGBR_truth_pred_err} for every type of simulation. The first row depicts the performance on the diagonal components of the pressure tensor, the second row on the off-diagonal components of the pressure tensor, and the final row on the heat flux vector.}
    \label{fig:R2-comparison-snapshot}
\end{figure}

To measure this discrepancy, the $R^2$ value is computed for this full reconnection region, for each simulation type and every predicted component. These results are displayed in Fig. \ref{fig:R2-comparison-snapshot}. These show that the MLP model has a consistently higher $R^2$ performance than the HGBR model for the diagonal pressure tensor components, independent of the simulation. On the remaining components, the HGBR in general still has the strongest performances. 

\subsection{Model evaluation along slices in the reconnection zone}

\begin{figure}[ht]
    \centering
    \includegraphics{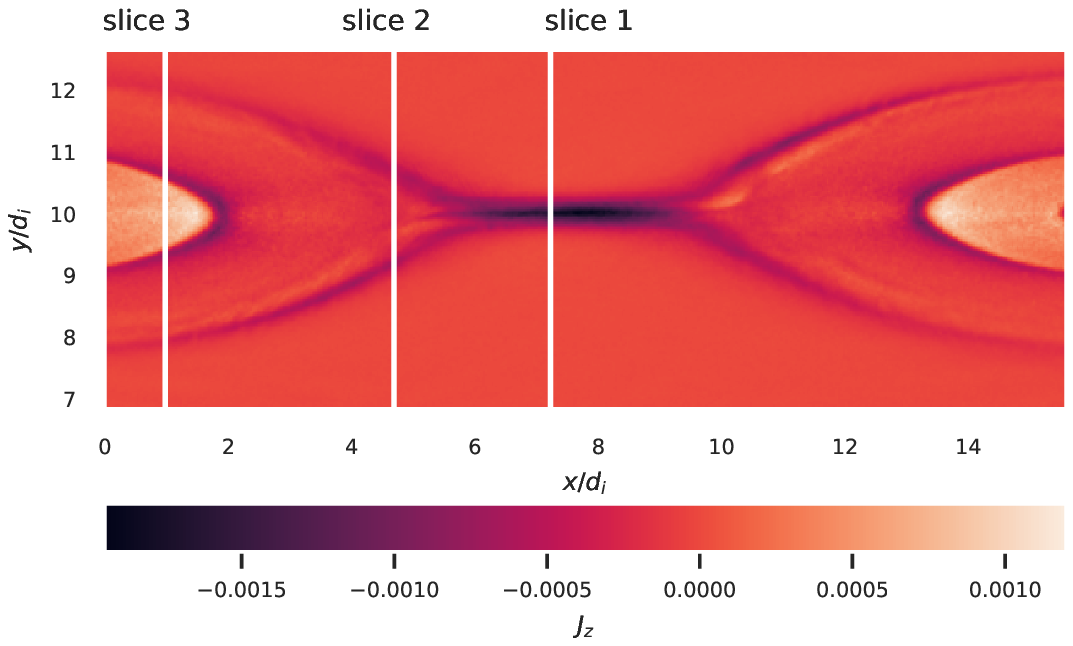}
    \caption{Slices of the reconnection area that are considered.}
    \label{fig:reconnectionslices}
\end{figure}

\begin{figure}
    \centering
    \includegraphics{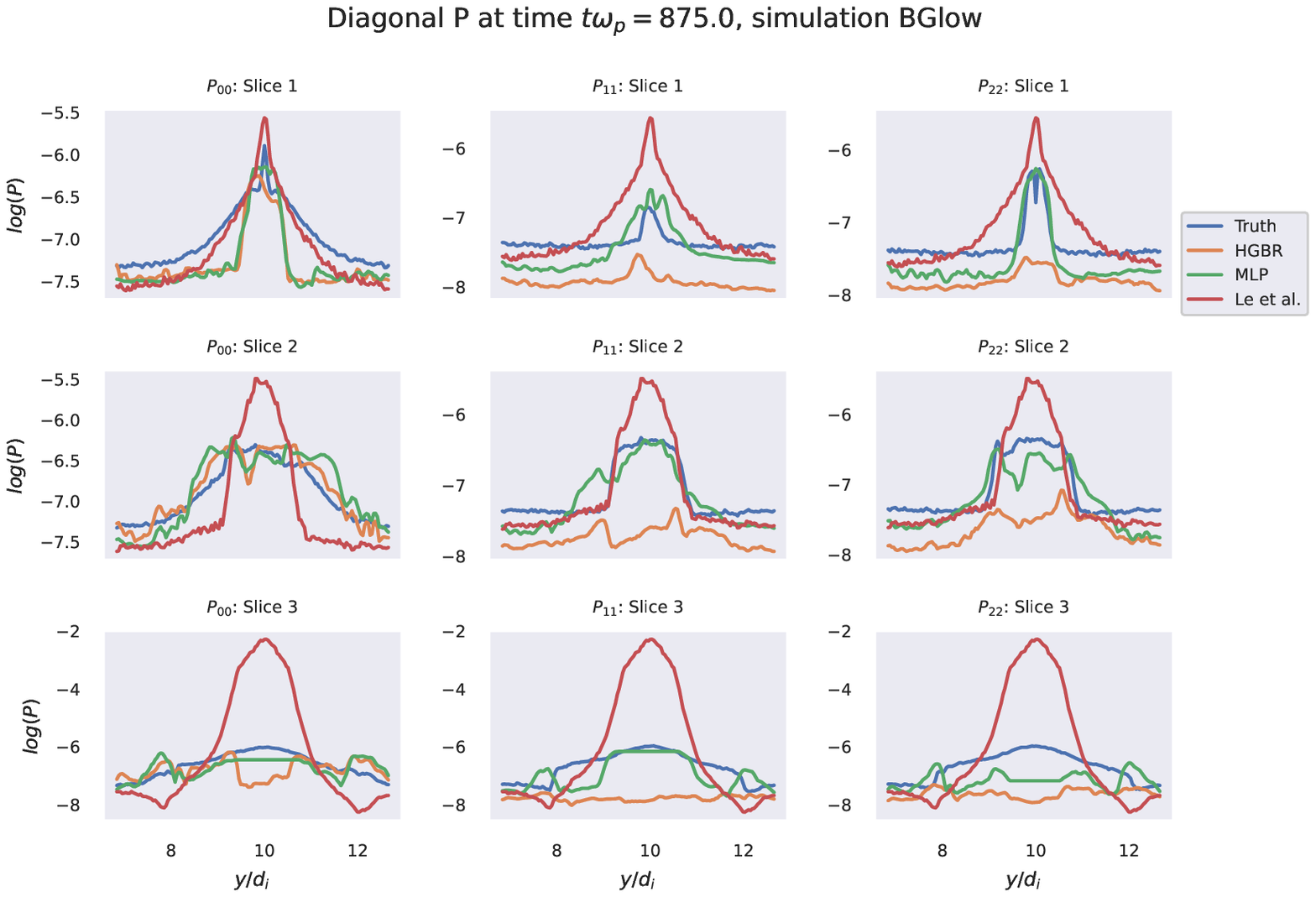}
    \caption{The prediction of the diagonal components of the pressure tensor along the slices shown in Fig. \ref{fig:reconnectionslices}. The simulation values are compared to the MLP model, the HGBR model and the closure of Le et al.}
    \label{fig:prediction_diagonal_comp_slices}
\end{figure}

\begin{figure}
    \centering
    \includegraphics{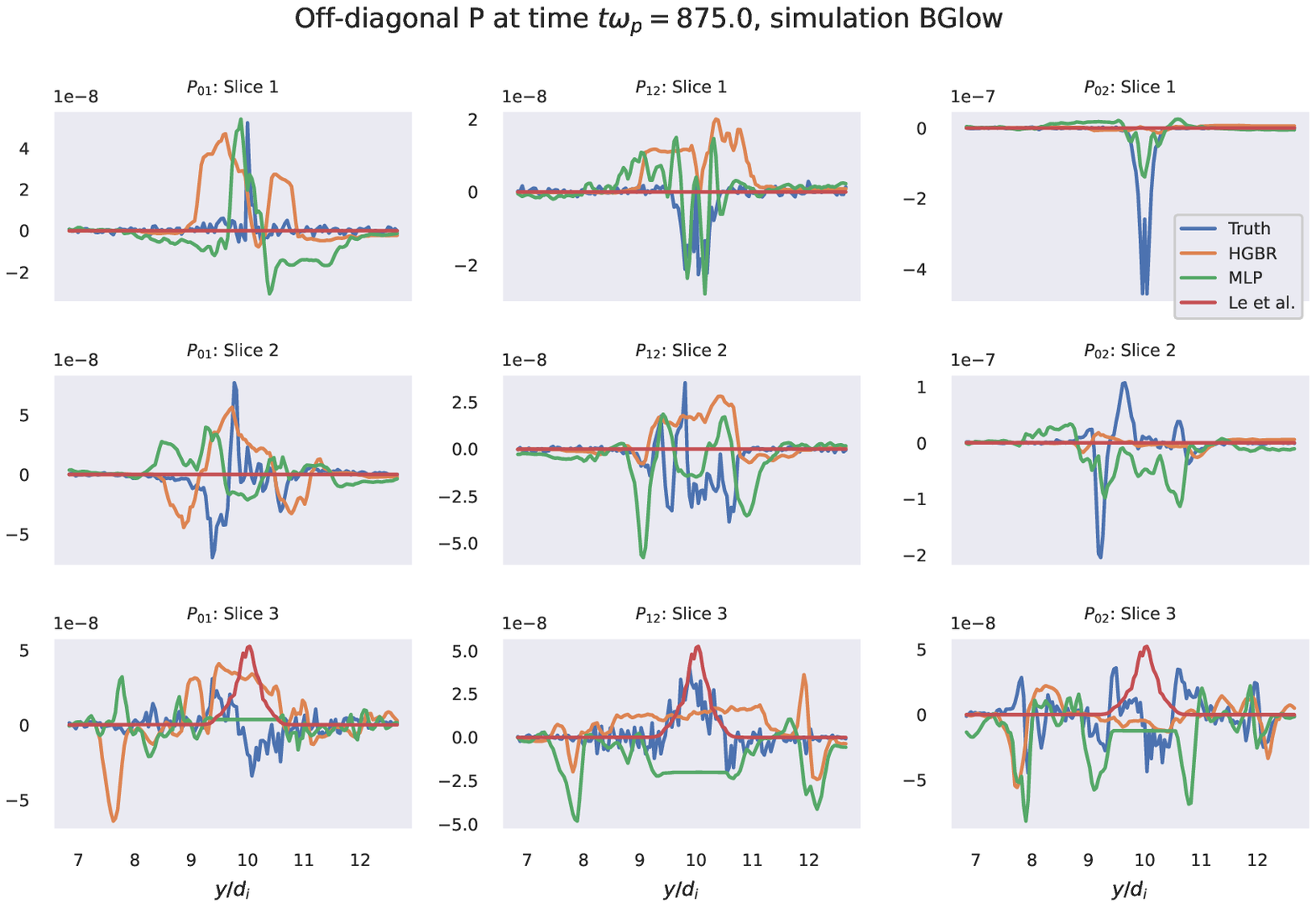}
    \caption{The prediction of the off-diagonal components of the pressure tensor along the slices shown in Fig. \ref{fig:reconnectionslices}. The simulation values are compared to the MLP model, the HGBR model and the closure of Le et al.}
    \label{fig:prediction_offdiagonal_comp_slices}
\end{figure}

\begin{figure}
    \centering
    \includegraphics{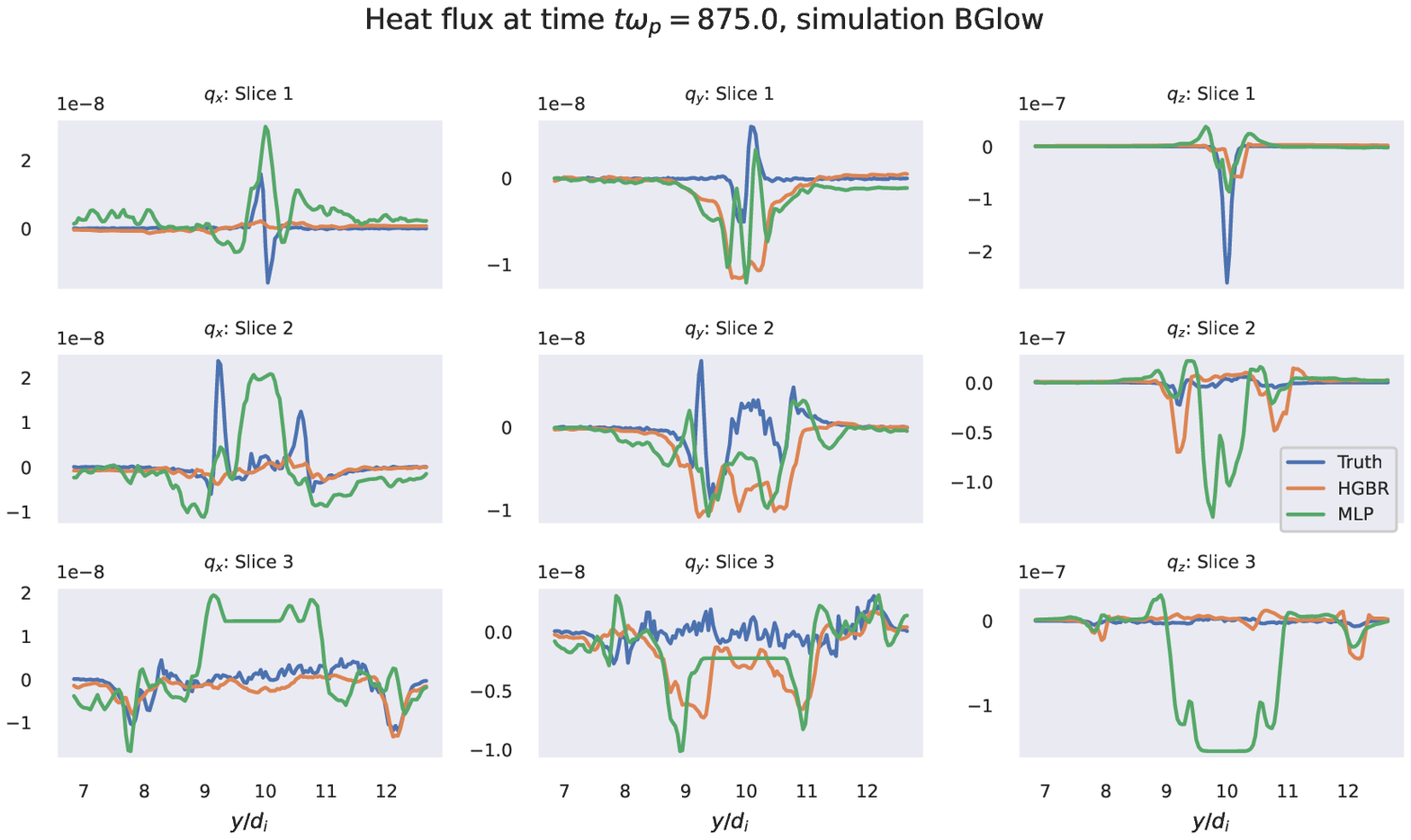}
    \caption{The prediction of the heat flux components along the slices shown in Fig. \ref{fig:reconnectionslices}. The simulation values are compared to the MLP model and the HGBR model}
    \label{fig:prediction_heatflux_comp_slices}
\end{figure}

Next, the models are evaluated along three slices in the reconnection zone. Once again, the snapshot of the simulation at time $t\omega_p = 875$ is used, taken  the region spanning $x \in [0, 15], y \in [7, 12]$, as shown in Fig. \ref{fig:reconnectionslices}. The linear regressor model is no longer considered in this experiment, and instead the models are compared to the the closure relation for the pressure tensor from Le et al., described in section \ref{chapter:fluidmodelsandmoments} by Eqs.~\eqref{eq:parallel_p_closure} - \eqref{eq:perp_p_closure}. The results for the diagonal pressure tensor components are shown in Fig \ref{fig:prediction_diagonal_comp_slices}, the off-diagonal components in Fig. \ref{fig:prediction_offdiagonal_comp_slices}, and the heat flux components in Fig. \ref{fig:prediction_heatflux_comp_slices}. We start with evaluating the models on the pressure tensor components, both diagonal and off-diagonal and end with the heat flux components. 

Observing the diagonal components of the pressure tensor in Fig. \ref{fig:prediction_diagonal_comp_slices}, we find that for the $\Ppar$ component, both ML models give a very accurate prediction of the peak observed when transitioning in and out of the reconnection zone. The MLP model gives the best prediction on the components $\Pperone$ and $\Ppertwo$ values, but does show a larger error on the X-lines themselves. This behavior could also be seen in the relative error shown in Fig. \ref{fig:MLP_truth_pred_err}.
Looking at the $\Pperone$ and $\Ppertwo$ components, the HGBR prediction is almost two orders of magnitude too small compared to the simulation. This confirms the observations of Fig. \ref{fig:HGBR_truth_pred_err}, where an underestimation was seen. The closure model by Le \textit{et al.} predicts the peak at the reconnection site accurately, but overestimates when going away from the reconnection zone. 

The off-diagonal components are shown in Fig. \ref{fig:prediction_offdiagonal_comp_slices}. These components are much more difficult to predict around the reconnection site, showing strong spatial variations. The model by Le \textit{et al.} is unable to capture these variations accurately, and always predicts a peak at the reconnection site. While both the ML models deviate from the zero line around the reconnection, indicating that something is happening, they are both unable to accurately capture the exact oscillations. Only the MLP model showed good results on slice 1, where the main peak was captured with good accuracy. This behavior was expected, since both Fig. \ref{fig:R2-comparison} and Fig \ref{fig:R2-comparison-snapshot} showed that the performance of the $R^2$ metric on the off-diagonal components was much weaker, indicating that the models have trouble with highly-varying values.

Finally, the prediction of the heat flux, shown in Fig. \ref{fig:prediction_heatflux_comp_slices} is studied. Once again, the same behavior is observed where strong spatial variations are seen in the heat flux inside or close to the reconnection region. The MLP model gives the best agreement along slice 1, accurately predicting the peaks and troughs along the reconnection region. However, in slice 2 and 3, the model predicts large troughs and peaks where there are none, giving high over-estimations of the actual heat flux value. 
The HGBR model can predict the most important peaks of the $q_x$ and $q_z$ component. Along the $q_y$ component, the model in general gives too low estimates of the heat flux inside the reconnection region. Notice that the HGBR model gives values close to zero outside the reconnection region, while the MLP model still predicts strong variations in that region. 

\section{Discussion}
\label{chapter:discussion}

First we determine if the ML models are better than the baseline model, the linear regressor. From the results shown in Fig. \ref{fig:R2-comparison}, we found that the ML models always had the strongest performance on the training sets. The performance on the test set was always better than the baseline model except for two components: the $P_{12}$ component of the pressure tensor and the $q_y$ component of the heat flux. These results can be explained by overfitting, since the performance on the training set of the ML models is very good, while on the test set it is outperformed by the linear regressor. We predict that a stronger fine-tuning of the training algorithm will result in a better performance of the ML models. For the remaining components, the ML models are the preferred models, showing clear improvements compared to the linear regressor. 

Next we compare the two ML models. From the performance shown in Fig. \ref{fig:R2-comparison}, we find that the HGBR model consistently has a better performance than the MLP model on the test set. From these results, we would determine that the HGBR model is the best model. However, when we made a visual observation of the model predictions on a reconnection region, as seen in Fig. \ref{fig:HGBR_truth_pred_err}, we observed that the HGBR model gave underestimations of the expected values. This is confirmed by the $R^2$ evaluation on the full reconnection region shown in Fig. \ref{fig:R2-comparison-snapshot}. These show that the MLP model consistently has a better performance than the HGBR model on all of the diagonal components of the pressure tensor, independent of the chosen simulation. Even though the HGBR model performed the best on the test set, its performance on the complete reconnection domain is less accurate than the MLP model.

We believe this discrepancy between evaluations is connected to how the test set is created. The data sets are created by sampling cells based on their agyrotropy. As described in section \ref{sect:create_datasets}, duplicate values are allowed in the training and validation set, to ensure the training algorithm focuses on all aspects of the simulation evenly. However, the test set does not contain duplicates, which causes an unbalance in the number of cells in- and outside of reconnection regions, with the number of cells in the reconnection region being in the minority. If the HGBR model performs significantly better on non-reconnection cells than the MLP model, then this could cause the HGBR model to perform better than the MLP model on the test set.

From the results seen in Fig. \ref{fig:MLP_truth_pred_err} and \ref{fig:HGBR_truth_pred_err}, we conclude that the MLP model is better in generalizing unknown values than the HGBR model. The MLP model provides an accurate reconstruction of the full reconnection zone, while the HGBR model can only accurately reconstruct the X-lines themselves, and not the region inside the reconnection zone, as observed in Fig. \ref{fig:prediction_diagonal_comp_slices}. Because we value the ability to generalize and interpolate highly for this type of problem, we choose the MLP model as our preferred model.

From the results shown in the previous Section, it is clear that the models had the best performance on the diagonal components of the pressure tensor. Both the off-diagonal components of the pressure tensor and the heat flux vector proved to be the most difficult to predict. We believe this is caused by three things. The first is the regional behavior of these components. The heat flux and the off-diagonal pressure tensor components are (close to) zero outside the reconnection region. Only inside the reconnection region do their values become important and start to fluctuate. This means that, in terms of the training set, a large portion of the sampled cells bring little to no contribution in the training of the models. The second reason is that the values show a noise-like behavior inside the reconnection region. On the X-lines, the models are able to give an accurate representation, as most often these show a peak in those components. However, in the regions inside the X-lines, these values become irregular and show a noise-like behavior, something clearly seen along slice 2 and 3 in Fig. \ref{fig:prediction_offdiagonal_comp_slices} and \ref{fig:prediction_heatflux_comp_slices}. It is unclear if how much of these fluctuations are caused by numerical noise from the simulation, or if these are actual physical values. 

We also would like to emphasize that the error inside the reconnection region is also in part caused by the sampling method. Fig. \ref{fig:agyro_dist}b) shows that the regions on the X-lines are well-represented in our sampling, something that is reflected well in Fig. \ref{fig:prediction_diagonal_comp_slices} - \ref{fig:prediction_heatflux_comp_slices}, while the regions inside the X-lines are not. A next step would be to create a better sampling method, that keeps the balanced data while also putting emphasis on the regions inside the X-lines.

It is also important to not forget that only local features are used to predict the targets. A next step is to include non-local information, and in the case of the heat flux, also information on the pressure, as an input. We believe that these new inputs would give a significant improvement to the predictions of the ML models, especially for the heat flux, because of the physical relations described in Section \ref{chapter:fluidmodelsandmoments}.

Finally, we would like to make a remark on the ML models as inherently interpolating machines. They learn their internal coefficients from the data that has been provided. If this data is broad, and covers multiple set-ups, it is possible to have a model that works well and consistent over the full problem domain. However, other cases have shown that ML models trained to emulate simple functions such as $f(x)=x$ can give unreliable results when tested outside the training domain\cite{gin2021deep, maulikNeuralNetworkRepresentability2020}. Since in our experiment, we are only considering four simulations, the models are expected to give larger errors when tested on data outside of this training set.
This brings the applicability of the ML model in question. It has to be trained on simulation data that is very expensive. If the model would be used as a surrogate model, we would need to provide it with a very broad data set of expensive kinetic simulations. However, once it has been trained on this data set, it can become an accurate surrogate model that has a much lower cost of use. 
While the ML model has not been trained on the complete problem domain, it is important to detect when the model is applied to problems that lie outside of its training domain. This detection is crucial in deciding if an error in a simulation is caused by the ML surrogate model being used outside its training domain, or another cause. Maulik et al.\cite{maulikNeuralNetworkRepresentability2020} discusses that one possible solution could be the creation of a non-linear embedding of the training and testing data into a two-dimensional space, a so-called T-distributed Stochastic Neighbor Embedding\cite{hinton2002stochastic}. By plotting the input features of both test and training set, it is possible to detect if an overlap exists of these two sets in the new embedded space. Another approach could be based on support vector machines\cite{suykens1999least} (SVM)'s. These map the input features with a kernel to a latent space, and can be clustered into regions with an SVM. By first forming clusters of the training data in the latent space, the test data can be checked if it falls into one of the created clusters. If not, the test data falls outside of the training domain.

\section{Summary and conclusion}

In this paper, two machine learning models are trained to learn an analytically unknown local closure relation for the full electron pressure tensor and the heat flux vector. The models are trained on four kinetic simulations of a Double Harris sheet simulation simulating magnetic reconnection, using the code iPiC3D. The four simulations have identical initial conditions except for the guiding background magnetic field, where four different values are selected. From the simulations, the first order moments and magnetic field are extracted and used as input features for the ML models.

The two machine learning models, a multi-layer perceptron and a gradient boosting algorithm, are first evaluated against a baseline model, a linear regressor. Their performance on the $R^2$ metric proofs that the ML models consistently perform better than the linear regressor on both training and test sets, and the use of the ML models is worth pursuing. 

The pressure tensor is divided in its diagonal and off-diagonal components. The ML models show comparable results with the empirical closure model of the pressure tensor created by Le et al.\cite{le2009equations} for both diagonal and off-diagonal components. 
Compared to each other, the MLP model shows better predictions than the HGBR model for the diagonal components on the complete reconnection region. It was shown that the MLP model can better generalize on new data, which we deem very important for this type of problem. For the off-diagonal components, both models show difficulty in predicting the many fluctuations of the components, especially inside the region bound by the reconnection's X-lines.

The heat flux vector proved to also be difficult to predict. At the reconnection site itself the MLP model is the best model. Further away from the reconnection site, the HGBR model gave the best predictions. 

The larger inside inside the reconnection region is partly caused by the sampling strategy used to construct the training and validation set of the models. The current strategy creates a strong emphasis on the X-lines themselves, but leaves the inside reconnection region under-sampled. A new sampling technique that provides both a balanced and more spread sampling will be a next step for this research. 

Both models are viable candidates for further experiments, but the context of the available data must be taken into account. The data comes from simulations, which are subject to numerical noise. The amount of simulations was also limited. Before the ML models can be used as surrogate models in numerical simulations, the training data must become more broad. A next step would be including more simulations with significant differences in the initial conditions. 

Finally, a local closure relation was assumed in these experiments. The inclusion of non-local information will be a next step in this research, and we expect this to have beneficial effects on the accuracy of the model. 

We conclude that the ML models show initial good results for finding a closure relation from kinetic simulations. Next steps would be determining a better sampling method, to ensure better results within the region bounded by the X-lines of the reconnection, a larger and broader training data set, and adding non-local features to the input.

\section*{Author's Contribution}
BL performed and analyzed the experiment and wrote the manuscript. GL created and provided the PiC simulation data and contributed to the introduction. GL and JA planned the study and GL, JA and SJ provided intellectual contribution to the manuscript. All authors took part in the manuscript revision and have read and approved the submitted version.

\section*{Acknowledgment}
This work was supported by the European Union's Horizon 2020 research and innovation programme under grant agreement No. 776262 (AIDA, Artificial Intelligence for Data Analysis, \url{www.aida-space.eu}) and the European Commission's H2020 program under the grant agreement EUHFORIA 2.0 (project n° 870405, \url{www.euhforia.com}).

\section*{Data Availability Statement}
The data that support the findings of this study are available from the corresponding author upon reasonable request.

\bibliography{bibliography}

\newpage

\appendix

\section{Configuration of the iPiC3D Simulation}
\label{appendix:simulation}
The size of the simulation box is chosen and fixed to $L_x \times L_y \times L_z =30 d_i \times 40 d_i \times 0.1 d_i$, where $d_i$ is the ion inertial length. Each spatial direction has respectively $769 \times 1025 \times 1$ cells with periodic boundary conditions in every spatial direction. 

Four species of particles are simulated, 2 electron species and 2 ion species. One set of ions and electrons are used as background as explained below, and the mass ratio between the ions and the electrons, $m_{i}/m_{e}$, is set to 256.

The electric field $\vec{E}$ is initialized to zero. The magnetic field $\vec{B}$ consists of a background component ($\vec{B}_0$) and a perturbation component ($\delta\vec{B}$), with the latter triggering the reconnection process. The magnetic field is defined as $\vec{B} = \vec{B}_0 + \delta \vec{B}$. The background part is initialized using a double hyperbolic tangent profile that switches the direction of the X component of the magnetic field:
\begin{align}
\begin{split}
B_x &= B_{x0} \left( \tanh \left(  \yBd \right) -  \tanh \left(  \yTd \right) -1.0 \right),  \\
B_y &= B_{y0}, \\
B_z &= B_{z0},  \label{eq:MagnFieldBase}
\end{split}
\end{align}
where $B_{x0}, B_{y0}$ and $B_{z0}$ are the constants, and $\delta$ is the thickness at half-height of the current layer, in this case set to $\delta = 0.5d_i$. 

The magnetic field perturbation is focalized at the midpoint of the X direction and at the location of each one of the two Harris layers in the Y direction with Gaussians:
\begin{align}
\begin{split}
    X_e &= \humpX, \\
    Y_{1/4} &= \humpB, \\
    Y_{3/4} &= \humpT, 
\end{split}
\end{align} 
These focal points are used to excite the magnetic field perturbation that triggers the emergence of reconnection. The following are the equations used to impose the perturbation:
\begin{align}
\begin{split}
\delta B_x &= 2 U_x B_{x0}  X_e \cdot \left( - \yBdy Y_{1/4} +  \yTdy Y_{3/4}  \right), \\
\delta B_y &= 2 U_x B_{x0} \xMdx X_e \cdot  \left(  Y_{1/4} - Y_{3/4}  \right), \\
\delta B_z &= 0.0,
\end{split} \label{eq:MagnFieldPert}
\end{align}
with $U_x= 0.4$, $\delta_x= 8 \delta$ and $\delta_y= 4 \delta$. 

\begin{figure}
    \centering
    \includegraphics[width=.5\textwidth]{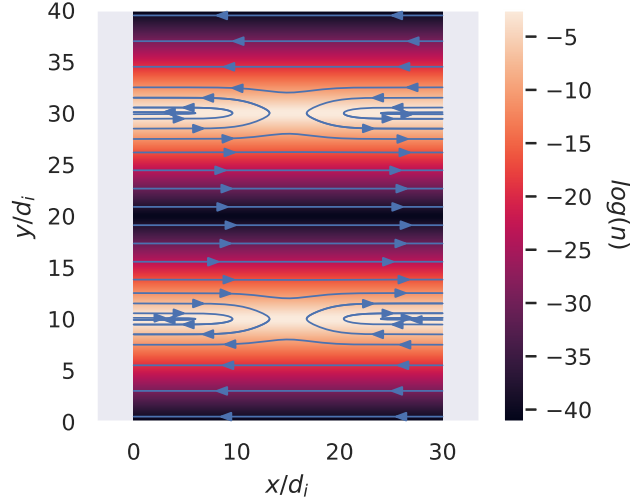}
    \caption{Configuration of the magnetic field described by equation \eqref{eq:MagnFieldBase} - \eqref{eq:MagnFieldPert} in the $x-y$ plane. We have taken $B_{x0}, B_{y0}, B_{z0} = 1$ for simplicity. The contours show the logarithm of the dimensionless particle density $n_d$.}
    \label{fig:MagnFieldConfig}
\end{figure}

The switch in direction imposed in the background magnetic field has to be balanced by a current, following Amp\`ere's law, for which one ion and one electron species are used to impose this charge neutral current. The following expression is derived from the application of Amp\`ere's law to the background magnetic field $\vec{B}_0$:

\begin{equation}
n_{d} = \dfrac{n_0}{4 \pi} \left(  \sech^2 \left( \yBd \right)  +  \sech^2 \left( \yTd \right)   \right),
\end{equation}

with $\sech$ the hyperbolic secant and $n_d$ the particle distribution of the background electron and ion species. The initial configuration of the magnetic field and particle distribution is shown in Fig. \ref{fig:MagnFieldConfig}.

The two remaining species, ion and electron, are to track the evolution of the reconnection process. After the particles and fields are initialized, multiple thousands of computational cycles update the location of the particles and the values of the magnetic and electric field. The PIC algorithm used in this work is explained in detail in Ref~\onlinecite{ricci2002simplified}. A single simulation is run for 20 000 cycles, where each cycle advances the simulation in time with a time step equal to the the inverse of the ion plasma frequency $\omega_{p,i}$: $\Delta t = 1/\omega_{p,i}= 0.0625$.

\section{Code units}
\label{appendix:codeunits}
Throughout the paper, all the quantities are normalized in code units with the following normalization factors. Here, $\tilde{X}$ is the quantity in code units:
\begin{align*}
    E &= \frac{c m_p \omega_{pp}}{e^-} \ \tilde{E} \\
    B &= \frac{m_p \omega_{pp}}{e^-} \ \tilde{B} \\
    J &= \frac{m_p c}{\mu_0 e^-} \ \tilde{J} \\
    V &= c \ \tilde{V} \\
    \rho &= N \ \tilde{\rho}
\end{align*}
with $c$ the speed of light in a vacuum in meter per second, $e^-$ the electron charge in Coulomb, $m_p$ the proton mass in kg, $\mu_0$ the vacuum permeability in Newton per square Ampere, $\omega_{pp}$ the proton plasma frequency in radials per second and $N$ a normalization factor with units Coulomb $m^{-3}$ that depends on the problem. 

\section{Creating training, validation test sets from the simulations}
\label{appendix:datasetconstruction}
From each of the four simulation, a training, validation and test set are constructed in the following steps:
\begin{enumerate}
    \item Extract 10 snapshots
    \item Assign each snapshot to either training, test or validation set (see Table \ref{table:datasetcreation})
    \item Sample cells from each snapshot based on the cell's agyrotropy.
    \item Extract the features described in section \ref{sect:data:input} from the sampled cells.
    \item Normalize training, validation and test set.
\end{enumerate}
At the end of this process, there are 12 data sets, consisting of a training, validation and test data set for each simulation. A training set contains, in total, \num{16000} cells, a validation set contains \num{4500} cells, and a test set contains \num{7500} cells.
Step 1 extracts 10 snapshots at fixed intervals from a simulation. Each snapshot contains all the field and particle information, frozen at the chosen time step. Each simulation is run for \num{20000} cycles, or time $t\omega_{p,i} = 1250$. The first snapshot is taken at cycle \num{10000} (to ensure reconnection has fully developed), and a new snapshot is taken after a \num{1000} computational cycles have passed. A single snapshot consists of 923MB of data, which gives 9.23GB of data extracted from each simulation, for a total of 36.92GB of data used to construct the training, testing and validation of the ML models.

In step 2 the snapshots are assigned to a specific data set as defined per Table \ref{table:datasetcreation}. This table displays which snapshots are assigned to either the training, validation and test set. Table \ref{table:guidingfield} is identical for each of the four simulations. 

\begin{table}[ht]
	\caption{The snapshots that are assigned to each data set. This is identical for each simulation.}
	\label{table:datasetcreation}
	\begin{ruledtabular}
	\begin{tabular}{cl}
		Data set & Snapshots \\
		\hline
		Training & 10000, 13000, 15000, 18000 \\
		Validation & 12000, 17000 \\
		Test  & 11000, 14000, 19000 \\
	\end{tabular}
	\end{ruledtabular}
\end{table}

\begin{figure*}
    \centering
    \includegraphics[width=.75\textwidth]{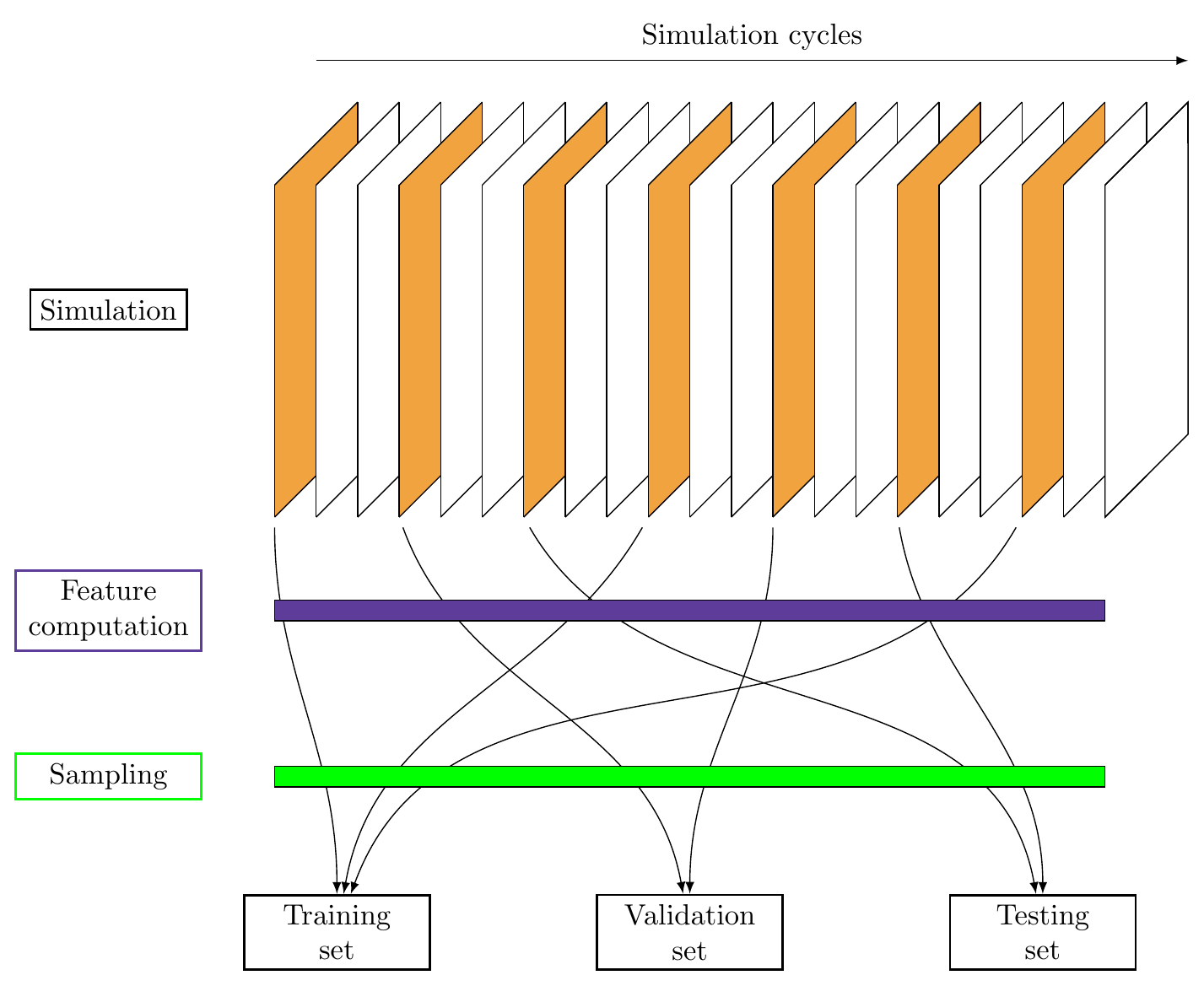}
    \caption{A simplified visualization of the process that creates a training, validation and test set are created from a simulation. Orange slices represent the snapshots, the purple bar represents the computation of the input features from the snapshot data and the green bar represents the cell-sampling based on the agyrotropy.}
    \label{fig:snapshots_to_data}
\end{figure*}

The third step samples cells from each snapshot based on the agyrotropy defined in Eq. \ref{eq:agyrotropy}. Inspection of the data, as shown in Figure \ref{fig:agyro_dist}, shows that the number of cells with interesting phenomena (such as reconnection) are far fewer than the number of cells where no relevant process is occurring at that time step. In order to ensure that the sampled data is not over-represented by the uninteresting cells, the agyrotropy is used to sample cells based on the influence of reconnection. The region from which the cells are sampled is also limited to the reconnection region itself, as shown in Fig. \ref{fig:agyro_dist}b).

After computing the agyrotropy of each cell, step 4 samples cells from each snapshot based on their agyrotropy. The agyrotropy distribution is split into a set of 5 bins. Because the distribution is right-skewed, the size of the bins has been based on a logarithmic distribution, to ensure an even coverage of all the values. The bins are visualized in Figure \ref{fig:agyro_dist} as vertical red lines. From each bin, a fixed number of cells are sampled. We have chosen to perform the sampling replacement, meaning that cells can be sampled multiple times from the same snapshot. This was done to ensure that each bin has the same number of points, as there are much fewer cells with a high agyrotropy. Because a snapshot is assigned to either a training, test or validation set, there is no possibility for leakage of data between the different sets because of the choice of sampling replacement. The number of cells extracted from a snapshot is decided by which data set the snapshot is assigned to. This is set to \num{4000}, \num{1500}, and \num{2500} cells from each sampling bin if the snapshot is assigned to the training, validation and test set, respectively. Thus a single simulation has a training set containing, in total, \num{16000} cells, a validation set containing \num{6000} cells, and a test set containing \num{10000} cells. This process is also displayed in Fig. \ref{fig:snapshots_to_data}.

In the final step, the constructed training, validation and test sets are normalized using various transformations. Normalizing has shown to give a positive effect on the stability and training speed of machine learning algorithms \citep{goodfellow2016deep}. We want to clarify that these transformations are only relevant for training the machine learning models. Once the model has been trained and provides predictions on the test sets, these predictions are transformed back to their native scale before any evaluation is done on them.

The input features are normalized with the min-max approach to the interval [0,1], by applying the following transformation to each feature:
\begin{equation}
\tilde{X}_{tr} = \frac{X_{tr} - \min_{tr}}{\max_{tr}-\min_{tr}}, \quad \tilde{X}_{val} = \frac{X_{val}  - \min_{tr}}{\max_{tr}-\min_{tr}}, \quad \tilde{X}_{test} = \frac{X_{test}  - \min_{tr}}{\max_{tr}-\min_{tr}}.
\end{equation}
Here $\max_{tr}$ and $\min_{tr}$ refer to the maximum and minimum value of each feature in the training set, respectively. The output targets are normalized by removing the mean and standard deviation from the data. This transformation is defined as
\begin{equation}
\tilde{Y}_{tr} = \frac{Y_{tr} - \mu_{tr}}{\sigma_{tr}}, \quad \tilde{Y}_{val} = \frac{Y_{val} - \mu_{tr}}{\sigma_{tr}}, \quad \tilde{Y}_{test} = \frac{Y_{test} - \mu_{tr}}{\sigma_{tr}}.
\label{eq:standardizing}
\end{equation}
Here $\mu_{tr}$ and $\sigma_{tr}$ are respectively the mean and standard deviation of the training set. The maximum, minimum, mean and standard deviation of the training data are used to prevent any information leakage \cite{camporealeChallengeMachineLearning2019}. Before removing the mean and standard deviation, the diagonal pressure components are transformed with a log scale transformation, which can be done because the pressure tensor components are
all strictly positive. 

\section{Hyperparameters of the ML models}
\label{appendix:hyperparameters}
In this section, the chosen hyperparameters of the ML models are provided. Both models only give a single feature as output. A total of 9 ML models and 9 HGBR models were trained, each with the same configuration.

\subsection{MLP configuration}

The following MLP configuration was used and constructed with PyTorch:
\begin{itemize}
    \item batch size: 400
    \item epochs: 100
    \item Optimizer: Adam 
    \item learning rate: $1.546 \cdot 10e-3$
    \item number of hidden layers: 4
    \item hidden layer 1: 80 neurons, Tanh activation function
    \item hidden layer 2: 110 neurons, Tanh activation function
    \item hidden layer 3: 150 neurons, ReLU activation function
    \item hidden layer 4: 80 neurons, Tanh activation function
\end{itemize}
The hidden layers are fully connected linear layers with an added activation function. The output layer is a fully connected linear layer without an activation function. 

\subsection{HGBR configuration}

The following HGBR configuration was used for the implementation in Scikit-Learn:
\begin{itemize}
    \item Loss function: least squares
    \item Learning rate: $5.878\cdot 10e-2$
    \item Max depth: 15
    \item Max iteration: 550
    \item Regularization: -4.3595
\end{itemize}

\end{document}